\documentclass[12pt]{article}

\usepackage[utf8]{inputenc}
\usepackage[T1]{fontenc}

\usepackage{setspace} 
\usepackage[vmargin = 1.5in, hmargin = 1.5in]{geometry}

\usepackage[usenames,dvipsnames,svgnames]{xcolor}
\usepackage{amsmath, amssymb, amsfonts, graphicx, tikz,pdflscape, mathtools, amsthm, upgreek, bm,pgfplots,csvsimple,multirow, multicol, booktabs}
\usepackage{verbatim}
\usepackage{natbib}
\usepackage{accents}

\usepackage{needspace}
\def\citeapos#1{\citeauthor{#1}'s (\citeyear{#1})}
\usepackage{comment}
\usepackage[inline]{enumitem}
\usepackage{tocvsec2}
\usepackage{threeparttable}
\usetikzlibrary{calc, cd}
\usepackage[skip = 10pt]{caption}
\allowdisplaybreaks[1]
\allowdisplaybreaks[1]

\definecolor{Blue}{RGB}{86,180,233}
\definecolor{Orange}{RGB}{230,159,0}
\definecolor{Green}{RGB}{0,158,115}
\definecolor{GmailBlue}{RGB}{42, 93, 176} 
\usepackage[
pagebackref,
colorlinks=true,
citecolor= GmailBlue,
linkcolor=GmailBlue,
urlcolor = GmailBlue
]{hyperref}

\newcommand{\bibtexorder}[1]{}

\usepackage{pgfplots}
\pgfplotsset{compat=newest}
\usepgfplotslibrary{groupplots}
\newcommand{\plotgap}{1.5 cm} 
\usepackage{tikz}
\usetikzlibrary{matrix,calc,shapes,arrows.meta,positioning}
\pgfplotsset{width = \textwidth/2}
\tikzstyle{hollow}=[circle,draw,inner sep=1.5]
\tikzstyle{solid}=[circle,draw,inner sep=1.5,fill=black]
\pgfplotsset{compat = newest}

\usepackage[capitalize,noabbrev]{cleveref}

\newtheoremstyle{breakital}
{}
{}
{\itshape}
{}
{\bfseries}
{}
{\newline}
{}

\theoremstyle{breakital}
\newtheorem{thm}{Theorem}
\newtheorem*{theorem*}{Theorem}
\newtheorem*{cor*}{Corollary}
\newtheorem{cor}{Corollary}
\newtheorem{prop}{Proposition}

\crefname{prop}{Proposition}{Propositions}
\crefname{thm}{Theorem}{Theorems}
\crefname{lem}{Lemma}{Lemmas}

\newtheoremstyle{break}
{}
{}
{}
{}
{\bfseries}
{}
{\newline}
{}

\theoremstyle{break}

\crefname{as}{Assumption}{Assumptions}

\theoremstyle{definition}

\newtheorem*{rem*}{Remark}

\numberwithin{lem2}{section}
\crefname{lem2}{Lemma}{Lemmas}

\def\a{\alpha}
\def\b{\beta}
\def\g{\gamma}
\def\d{\delta}
\def\e{\varepsilon}

\def\h{\eta}
\def\th{\theta}

\def\k{\kappa}
\def\l{\lambda}

\def\s{\sigma}
\def\t{\uptau}

\def\w{\omega}

\label{key}

\def\D{\Delta}

\def\L{\Lambda}

\def\W{\Omega}

\def\R{\mathbf{R}}

\def\CC{\mathcal{C}}

\def\FF{\mathcal{F}}
\def\GG{\mathcal{G}}
\def\HH{\mathcal{H}}

\def\SS{\mathcal{S}}

\def\VV{\mathcal{V}}

\def\P{\mathbf{P}}

\DeclareMathOperator{\E}{\mathbf{E}}
\DeclareMathOperator*{\argmax}{argmax}

\DeclareMathOperator{\Var}{var} 
\DeclareMathOperator{\var}{var}

\newcommand{\de}{\mathop{}\!\mathrm{d}}

\newcommand{\Paren}[1]{\left( #1 \right)}

\newcommand{\brac}[1]{[ #1 ]} 

\newcommand{\Brac}[1]{\left[ #1 \right]}

\newcommand{\Set}[1]{\left\{ #1 \right\}}

\title{Dynamic Information Provision:\\ Rewarding the Past and Guiding the Future}
\author{Ian Ball\thanks{
		Department of Economics, MIT, \texttt{ianball@mit.edu}. 
		I thank Dirk Bergemann for thoughtful and friendly guidance throughout the project. An earlier version of this paper appeared as the second chapter of my Ph.D.\ dissertation at Yale. For helpful discussions, I thank Alexandre Belloni, Ben Brooks, Eduardo Faingold, Mira Frick, Drew Fudenberg, Simone Galperti, Marina Halac, Johannes H\"{o}rner, Ryota Ijima, Navin Kartik, Emir Kamenica, Deniz Kattwinkel, Nicolas Lambert, Chiara Margaria, Stephen Morris, Giuseppe Moscarini, Larry Samuelson, Andy Skrzypacz, Philipp Strack, and Alex Wolitzky. I thank audiences at Yale, Allerton, Stony Brook, and the Stanford Summer Institute in Theoretical Economics (SITE) for useful feedback. Finally, I thank Yifan Dai for excellent research assistance. 
	}
}

\date{16 March 2023}

\begin{document}
	 
\maketitle

\begin{abstract}
I study the optimal provision of information in a long-term relationship between a sender and a receiver. The sender observes a persistent, evolving state and commits to send signals over time to the receiver, who sequentially chooses public actions that affect the welfare of both players. I solve for the sender's optimal policy in closed form: the sender reports the value of the state with a delay that shrinks over time and eventually vanishes. Even when the receiver knows the current state, the sender retains leverage by threatening to conceal the future evolution of the state. 



\end{abstract}


\newpage

\section{Introduction} \label{sec:introduction}

How can information be used in place of money as a reward to motivate behavior? In many relationships, controlling information provides powerful leverage. A leaker sharing protected information with a media outlet can use the promise of additional leaks to demand slanted coverage. Within organizations, from large firms to the military, turf wars impede efficient information exchange \citep{HerreraEtal2017}. In such organizations, where preferences are misaligned and transfers are often infeasible, refusing to share information can be an important bargaining chip. 
 
In these relationships, between individuals or organizational units, one party controls information that is necessary to guide efficient decisions. When preferences are misaligned, this information can be withheld as a punishment for past behavior. In this paper, I illustrate how the dual role of information shapes the dynamics of optimal information provision within a long-term relationship.

I consider a stylized model in continuous time with a sender (she) and a receiver (he). There is a payoff-relevant state, which follows a diffusion process. The sender observes the evolution of this state, and she sends signals over time to the receiver, who sequentially chooses public actions that affect the welfare of both players. The sender and receiver have partially aligned quadratic preferences. The receiver wants to match his action with the state, but the sender wants the receiver to shift his action above the state by a fixed bias. 

The only instrument available to the sender is the control of information. The sender commits to a dynamic information policy, which assigns a signal distribution to each private history of past states, signals, and actions. Over time, as the sender observes the evolving state and the receiver's actions, she sends the signals prescribed by the information policy. The receiver observes the sender's signals and chooses actions sequentially. Since the receiver is forward-looking, he considers the effect of his actions both on his flow payoff and on the informativeness of future signals. The sender, anticipating the receiver's best response, chooses an information policy to maximize her expected utility.

The fundamental tradeoff for the sender is between \emph{precision} and \emph{bias}. Providing the receiver with information about the current state has two effects. On the one hand, the receiver can more precisely tailor his current action to the state, making both players better off. On the other hand, since the state process is persistent, the receiver's uncertainty about future states is reduced, making future information less valuable to him. Hence, the receiver is less willing to bias his actions. 

The sender faces a complex, nested optimization problem. I reduce the sender's problem in two steps. 

First, I change the domain of optimization from information policies to \emph{decision rules}, i.e., state-dependent distributions of actions over time.  A decision rule is a best response to some information policy if and only if it is a best response to a canonical information policy---the associated direct, grim-trigger policy that makes direct action recommendations  and cuts off all future information if any recommendation is violated. Decision rules that can be induced in this way are called \emph{obedient}. 

Second, I observe that the players' payoffs and the obedience constraint can be expressed in terms of two statistics of a decision rule---the bias and variance. At a given time, the bias%
\footnote{This (action) bias is a statistic computed from a decision rule. The (preference) bias is a parameter of the sender's utility function.} is the difference between the action recommended to the receiver and the receiver's bliss point (his expectation of the state). The variance is the posterior variance of the current state, given the history of action recommendations. The bias and variance are stochastic processes, but I show that  deterministic bias and variance processes---termed paths---trace out the Pareto frontier. 

Bayesian updating restricts which variance paths are feasible. For a variance path to be induced by some information policy, a necessary condition called Bayes plausibility is that the variance never increases by more than it would in the absence of new information. I show that this condition is also sufficient. \cref{thm:delayed_reporting} says that every Bayes-plausible posterior variance path can be induced by \textit{delayed reporting}. At each time $t$, the sender's action recommendation reveals the state realization at some earlier time $\varphi(t)$.

The sender's problem is therefore reduced to optimizing over obedient, Bayes-plausible bias--variance paths. I use Lagrangian relaxation and dynamic programming to find the solution in closed form. The optimal bias and variance functions are stated in \cref{thm:optimal_policy}. There are two cases. 

If the sender's bias is sufficiently small relative to the volatility of the state process, then the sender can induce her first-best decision rule. She keeps the receiver perfectly informed of the state but threatens to cut off all future information if the receiver ever deviates from the \emph{sender's} optimal action. 

If the sender's bias is sufficiently large relative to the volatility of the state process, then the sender cannot induce her first-best decision rule. Hence, the bias--precision tradeoff is in force. The optimal policy has two phases. First, in the \emph{transition phase}, the sender gradually reduces the receiver's uncertainty about the current state  while gradually narrowing the gap between the recommended action and the receiver's bliss point. At some finite time, the current state is fully revealed, and the \emph{stationary phase} begins.  Thereafter, the sender keeps the receiver perfectly informed of the current state. The sender demands that the receiver maintain a fixed bias between his action and the state. If the receiver deviates at any time, the sender cuts off all future information. 

The rest of the paper is organized as follows.  \cref{sec:related_literature} reviews related literature. \cref{sec:model} presents the model.  \cref{sec:two_period} solves a two-period example. The main analysis begins in \cref{sec:deterministic}, where I introduce delayed reporting and simplify the sender's problem. \cref{sec:solution} describes the optimal information policy and analyzes comparative statics. \cref{sec:multidim} solves for the optimal policy with multidimensional states and actions. The components of the state are revealed sequentially, in order of increasing persistence. \cref{sec:conclusion} is the conclusion.  Measure-theoretic definitions are in \cref{sec:strategies}. Proofs are in \cref{sec:proofs}.

\subsection{Related literature} \label{sec:related_literature}

My paper joins a growing literature on dynamic information design. The first papers \citep{RenaultSolanVieille2017,Ely2017} study the optimal provision of information (about an evolving state) to a myopic receiver who acts repeatedly.\footnote{The working paper version \cite{Ely2015wp} finds that the solution is unchanged if the receiver is forward-looking. This is a special feature of the binary-action setting.} Since the receiver is myopic, his action choices depend only on his beliefs about the current state, not the promise of future information.

The closest dynamic information design papers \citep{Smolin2017wp,ElySzydlowski2018fc,OrlovSkrzypaczZryumov2018wp} study the optimal provision of information over time to a forward-looking receiver who chooses when to stop.\footnote{Other dynamic information design papers explore different issues: costly communication \citep{Honryo2018}; private information held by the receiver \citep{Au2015}; and the effect of a decision deadline \citep{BizottoRudigerVigier2017wp}.} To stop optimally, the receiver considers his current belief about the state and also the future information he will receive if he waits. The sender can delay information transmission in order to entice the receiver to wait longer. In my paper, by contrast, the receiver chooses a rich action at each time. Therefore, the sender has a countervailing motive to reveal information so that the receiver can adjust his action to the current state.\footnote{In subsequent work by \cite{Kaya2022wp}, the receiver chooses effort each period. But there is no precision motive because the sender prefers more effort, no matter the state.} This precision motive drives gradual information revelation.\footnote{ \cite{OrlovSkrzypaczZryumov2018wp} find an equilibrium with gradual information revelation in a model in which the sender has only within-period commitment. If the sender's  bias against stopping is sufficiently strong, then promising delayed information is not credible. In my model, gradualism arises with or without intertemporal commitment; see the discussion of commitment in \cref{sec:discussion}.}

The dynamics in my model are broadly similar to the backloading of rewards in dynamic principal-agent models with cash constraints \citep{Lazear1981,HarrisHolmstrom1982,ThomasWorrall1994}.\footnote{In a quite general (complete information) principal--agent setting, \cite{Ray2002} shows that in all efficient ``self-enforcing agreements,'' the continuation value of the agent is backloaded.} Withholding information, unlike money, directly entails inefficiency. Closest to my paper is the apprenticeship model in the independent work of  \citet[][hereafter FR]{FudenbergRayo2019}, which builds upon the framework of \cite{GaricanoRayo2017}. In FR, the principal is endowed with a perfectly divisible unit of knowledge that can be costlessly transmitted to the agent, who is cash-constrained. The principal selects a contract specifying paths of knowledge, effort, and wages for the agent. At any time, the agent can walk away with his current knowledge stock. While immediate knowledge transmission is efficient, the principal's optimal contract transmits knowledge gradually in order to suppress the value of the agent's outside option. 

In my model, the receiver's posterior variance and action bias, respectively, play similar roles to the principal's stock of untransmitted knowledge and the agent's excess effort in FR. But there are important differences. First, in my model the information provided by the sender grows stale over time (because the state evolves). So even if the receiver knows the current state, it is still \emph{feasible} for the sender to demand bias from the receiver. Second, the players' payoffs in my model are not transferable. The curvature in the Pareto frontier makes it \emph{optimal} for the sender to demand bias after fully revealing the state. The stationary phase, featuring perfect information transmission and biased actions, is new to my model. 

\section{Model} \label{sec:model}

Time is continuous and the horizon is infinite. There are two players: a sender (she) and a receiver (he). At each time $t \in [0, \infty)$, the state $\th_t \in \R$ is realized and the receiver chooses an action $a_t \in \R$. Flow utilities for the sender and receiver are given by
\begin{equation*}
	u_S (a_t, \th_t) = - (a_t - \th_t - \b)^2, 
	\qquad 
	u_R (a_t, \th_t) = - (a_t - \th_t)^2.
\end{equation*}
The receiver wants to match his action with the state, but the sender wants the receiver to shift his action above the state by a bias $\b$. Without loss, assume $\b > 0$. Both players discount future flow utilities at exponential rate $r$. 

The initial state $\th_0$ is normally distributed with mean $\mu_0$ and variance $\s_0^2$. The state then follows the stochastic differential equation
\begin{equation} \label{eq:SDE}
		\de \th_t = \k \th_t \de t + \s \de Z_t,
\end{equation}
where $\{ Z_t \}_{t \geq 0 }$ is a standard Brownian motion, independent of the initial state $\th_0$.%
\footnote{The process is normalized to have zero drift if $\th_t = 0$. The results do not change if  \eqref{eq:SDE} is replaced with $\de \th_t = (\mu_t + \kappa \th_t) \de t + \sigma \de Z_t$, for any deterministic time-dependent drift $\mu_t$.} The volatility parameter $\s$ is strictly positive. Assume $2 \k < r$. Hence, a constant action yields finite expected utility for both players. The process can be explosive ($\k > 0$), mean-reverting $(\k  <0$), or a Brownian motion ($\k = 0$). 
 
The state distribution is common knowledge, but the state realizations are observed only by the sender. The sender also observes the receiver's actions.  Initially, the sender commits to a dynamic \emph{information policy} $S$, which consists of a signal realization space $\mathbf{S}$ together with a signal distribution for each history of past states, signals, and actions. Given this policy, the receiver faces a dynamic decision problem. At each time $t$, having observed the signals sent up to time $t$, the receiver chooses an action.\footnote{Restricting to pure strategies is without loss. Any randomization by the receiver can be  incorporated into the sender's signal.} The receiver learns about the state only from the sender's signals. In particular, the receiver does not observe past flow payoffs.

\subsection{Sender's problem}

The sender's information policy and the receiver's action choices jointly determine a stochastic process $A = \{ A_t\}_{t \geq 0}$, which I called a \emph{decision rule} as in \cite{BergemannMorris2016}. A decision rule $A$ yields utilities
\[
	u_i (A) = \E \Brac{ \int_{0}^{\infty} r e^{-r t} u_i (A_t, \th_t)}, \qquad i = S,R.
\]
A decision rule $A$ is \emph{compatible} with an information policy $S$ if at each time $t$, the sender's signals up to time $t$ (which depend on the receiver's previous actions $A_s$ for $s < t$) provide sufficient information for the receiver to select action $A_t$.\footnote{This circularity makes the formal definition delicate; see \cref{sec:strategies}.} A decision rule $A$ is a best response to an information policy $S$ if $A$ maximizes the receiver's utility over all decision rules compatible with $S$. The sender  maximizes her utility $u_S(A)$ over all pairs $(S,A)$ with the property that $A$ is a best response to $S$. 

\subsection{Discussion of assumptions} \label{sec:discussion}

\paragraph{Monitoring} The sender perfectly observes the receiver's actions. Therefore, the sender can condition future information on past actions. Without monitoring, the receiver would act myopically, always matching his action with his expectation of the state. The sender, in turn, would perfectly reveal the state at all times. 

\paragraph{Commitment}
The sender is assumed to have full dynamic commitment power. That is, the sender can commit to action-contingent signals within each ``period'' and across ``periods.''\footnote{This distinction can be formalized in a discrete-time approximation of the continuous-time model.}
One-period commitment is crucial, as in static Bayesian persuasion \citep{KamenicaGentzkow2011}. Multi-period commitment is not necessary in the main case of interest.\footnote{That is, if  $\b \geq \s / \sqrt{r -2 \k}$. Otherwise, the receiver is promised more than his reservation utility, so one-period punishments are insufficient to motivate the receiver. Therefore, multi-period commitment is needed.} Under the optimal full-commitment policy, the receiver is never promised more than his reservation utility. Therefore,  the sender can achieve the same decision rule with an alternative policy that uses only one-period punishments.  If the receiver deviates, the sender provides no information next period. Thereafter, the sender reverts to the policy that is optimal given the receiver's current beliefs. This reversion does not increase the receiver's continuation payoff, so the receiver's payoff from deviating is unchanged. 

\paragraph{Unobserved payoffs} The receiver does not observe his own flow payoffs. With quadratic utility, flow payoffs perfectly reveal the trajectory of the state, so this assumption is necessary to preserve the receiver's uncertainty. In the motivating applications, it seems reasonable that the receiver could experience the cost of his own state uncertainty without learning the state realization. In any case, the forces in the model should apply as long as the sender retains some informational advantage over the receiver. 
 
\section{Two-period example} \label{sec:two_period}

This section illustrates the sender's bias--precision tradeoff in an example with two periods, $t = 1,2$. Flow payoffs are as in the main model. Each player $i$
maximizes the discounted sum of flow payoffs $u_{i,1} + \d u_{i,2}$. The state process is given by
\[
	\th_1 \sim N(0, \s_1^2), \qquad \th_2 = \rho \th_1 + \e, \qquad \e \sim N(0, \s^2),
\]
where $\e$ is independent of $\th_1$. 

This example can be analyzed backwards. In the second period, the receiver acts myopically since this is the last period of the game.  Full disclosure maximizes flow payoffs for both players, so it is optimal for the sender to fully disclose the state on-path, i.e., if the receiver obeys the sender's first-period recommendation. If the receiver disobeys the first-period recommendation, it is optimal to impose the maximal punishment by providing no information.

Now consider the first period. If the sender's signal induces posterior variance $v_1$, the sender can demand that the receiver bias his action away from his posterior mean by $b_1$, provided that
\begin{equation} \label{eq:bias_variance_tradeoff}
	b_1^2 \leq \d (\rho^2 v_1 + \s^2).
\end{equation}
This inequality captures the bias--precision tradeoff. The right side---the discounted residual variance of $\th_2$ without additional information---is the receiver's discounted loss from not learning the state in the second period. When this constraint binds, the sender must pay a price of $1/(\d \rho^2)$ in higher variance per unit of squared bias. 

The sender chooses $b_1$ and $v_1$ to maximize her total (on-path) discounted payoff $- (b_1 - \b)^2 - v_1 - \d \b^2$, subject to \eqref{eq:bias_variance_tradeoff}. There are three cases.

\begin{enumerate}[label = (\roman*), ref = \roman*]
	\item \label{it:not_binding} $\b \leq \sqrt{\d} \s$. Here, \eqref{eq:bias_variance_tradeoff} is not binding, so the optimum is $\hat{b}_1 = \b$ and $\hat{v}_1 = 0$. 
	\item \label{it:rho} $\rho = 0$. Here, $v_1$ does not appear in \eqref{eq:bias_variance_tradeoff}. Revealing $\th_1$ does not affect the receiver's uncertainty about $\th_2$, so there is no tradeoff between bias and precision. The optimum is  $\hat{b}_1 = \min \{ \b, \sqrt{\d} \s \}$ and $\hat{v}_1 = 0$.
	\item \label{it:binding} $\b > \sqrt{\d} \s$ and $\rho \neq 0$. This is the main case of interest. Here, \eqref{eq:bias_variance_tradeoff} must hold with equality; otherwise, the sender would strictly prefer to reveal more in the first period. The first-order condition for the optimal bias equates the direct marginal benefit $2 (\b - b_1)$ of increasing $b_1$ with the marginal cost $2 b_1 / (\d \rho^2)$ from the required increase in variance.\footnote{The solution is given by 
		\[
		\hat{b}_1 = \b \frac{ \d \rho^2}{1 + \d \rho^2}, \qquad \hat{v}_1 = \b^2 \frac{ \d \rho^2}{(1 + \d \rho^2)^2} - \frac{\s^2}{\rho^2}.
		\]
		As a function of $v_1$, the sender's payoff is concave, so it is optimal to induce a constant variance. Here I assume that $\s_1^2 \geq \hat{v}_1$, so there exists a signal structure inducing this variance $\hat{v}_1$.}
		
		I now describe in detail the comparative statics for the optimal bias $\hat{b}_1$ and variance $\hat{v}_1$ since the same results hold for the continuous-time solution during the transition phase. As the preference bias $\b$ increases, the sender withholds more information (higher $\hat{v}_1$) in order to induce greater action bias (higher $\hat{b}_1$). As the volatility $\s$ increases, the bias $\hat{b}_1$ is unchanged (since $\s$ does  not enter the bias first-order condition), but the variance $\hat{v}_1$ decreases since less variance is needed to induce the same level of bias. As $\d$ and $\rho^2$ increase, the receiver's loss from not learning $\th_2$ is more sensitive to the first-period variance $v_1$. That is, the price $1/ (\d \rho^2)$ of bias (in units of variance) decreases.  Hence, $\hat{b}_1$ increases. The effect on total ``spending'' $\hat{v}_1$ is ambiguous. 
\end{enumerate}

\section{Reducing the dimension of the problem} \label{sec:deterministic}

In this section, I change the domain of the sender's optimization problem from the space of all information policies to the simpler space of obedient decision rules with deterministic bias and variance. 

\subsection{Obedient decision rules}

For a given information policy, the sender's utility depends only on the decision rule induced by the receiver's best response. Therefore, it suffices to optimize over every decision rule that can be induced as a best response to some information policy. I now characterize such decision rules. 

Given a decision rule $A$, let $\FF_t^A$ denote the $\s$-algebra generated by $A_s$ for $s \leq t$. This $\s$-algebra represents the minimal information that the receiver must have by time $t$ in order to follow $A$. The receiver gets exactly this minimal information if the sender makes direct action recommendations. 

Decision rule $A$ is \emph{obedient} if for each $t \geq 0$,\footnote{All inequalities involving conditional expectations are interpreted almost surely.}
\begin{equation} \label{eq:obedience}
\begin{aligned}
- \E \Brac{ \int_{t}^{\infty} r e^{-r (s-t)} \Paren{ A_s - \th_s}^2 \de s  \Big\vert \FF_t^A}
\geq - \frac{\s^2 + r \Var(\th_t | \FF_t^A)}{r - 2 \k}.
\end{aligned}
\end{equation}
Inequality \eqref{eq:obedience} is the \emph{time-$t$ obedience constraint}. Conditional on the history of actions up to time $t$, 
the receiver's expected continuation value from following the decision rule is at least as large as his reservation utility---his expected continuation value from deviating at time $t$, forfeiting all future information, and choosing $A_s = \E [ \th_s | \FF_t^A] = e^{\k (s - t)} \E [ \th_t | \FF_t^A]$ for $s \geq t$.

\begin{prop}[Obedience] \label{res:obedience} Decision rule $A$ is a best response to some information policy if and only if $A$ is obedient. 
\end{prop}

The necessity of obedience is clear. If obedience is violated, then at some time $t$ the receiver can profit by deviating to myopic play on a nontrivial subset of action histories. For sufficiency, I show than any obedient decision rule $A$ is a best response to a canonical information policy---the direct, grim-trigger information policy that sends signals $S_t = A_t$, provided that the receiver has followed all past recommendations. If the receiver ever deviates, this policy sends uninformative signals forever after. The time-$t$ obedience constraint ensures that it is not profitable for the receiver to start deviating at time $t$. 

The sender's problem is now reduced to maximizing over all obedient decision rules. I next  show that it suffices to maximize over simpler summary statistics of these decisions rules. 

\subsection{Bias and variance} \label{sec:delayed_reporting}

For any decision rule $A$, define the induced bias and variance processes by
\[
	B_t = A_t - \E [\th_t | \FF_t^A]
	\quad \text{and} \quad
	V_t = \Var(\th_t | \FF_t^A).
\]
The \emph{bias} $B_t$ is the gap at time $t$ between the action and the receiver's expectation of the state, given the action recommendations up until time $t$. The \emph{variance} $V_t$ is the posterior variance of the state $\th_t$, given the action recommendations up until time $t$. The players' payoffs and the obedience condition \eqref{eq:obedience} can be expressed in terms of the bias and variance processes only.\footnote{In particular, the expected state $\E[ \th_t | \FF_t^A]$ does not appear by itself because the sender's bias $\b$ is state-independent.} In general, $B_t$ and $V_t$ are random, but I show that the Pareto frontier is traced out by decision rules with deterministic bias and variance. 

\begin{prop}[Deterministic bias and variance] \label{res:bias_variance}
	For each obedient decision rule $A$, there exists an obedient decision rule $A'$ with deterministic bias and variance such that  both players weakly prefer $A'$ to $A$.
\end{prop}

Here is a sketch of the proof. Consider an obedient decision rule $A$ with bias and variance processes $\{ B_t\}_{t \geq 0}$ and $\{ V_t\}_{t \geq 0}$. Let $b(t) = \E [B_t]$ and $v(t) = \E [V_t]$ for each $t$.  I will construct a new decision rule $A'$ with deterministic bias $b$ and deterministic variance $v$. Both players weakly prefer $A'$ to $A$ because their flow payoffs are linear in variance and strictly concave in bias. The receiver's reservation utility is linear in variance, so $A'$ is obedient. To complete the proof of \cref{res:bias_variance}, I need to construct a signal structure that induces the desired variance function $v$. The next section introduces a convenient general construction called delayed reporting. 

\subsection{Bayes plausibility and delayed reporting}

First I introduce a necessary condition for a variance process to be consistent with Bayesian updating. Suppose that at time $t$, the posterior variance of $\th_t$ is $V_t$. If no additional information is provided, then at time $t + h$, the posterior variance of $\th_{t + h}$ is $\h( V_t, h)$, where the function $\h$ is defined by
\begin{equation} \label{eq:var_exp}
	\h(v,h) =
	\begin{cases}
	 e^{2 \k h} v + (e^{2 \k h} - 1) \frac{\s^2}{2 \k} &\text{if}~ \k \neq 0, \\
	  v + \s^2 h &\text{if}~ \k = 0.
	  \end{cases}
\end{equation}
 A function $v \colon [0,\infty) \to [0, \infty)$ is \emph{Bayes-plausible} if
\begin{enumerate}[label  = (\roman*), ref = \roman*]
	\item  \label{it:initial_variance} $v(0) \leq \s_0^2$;
	\item \label{it:dynamic_variance} $	v(s) \leq \h(v(t), s - t)$ for all $s > t \geq 0$.
\end{enumerate}
By the law of total variance, a necessary condition for a variance process $\{ V_t\}_{t \geq 0}$ to be induced by some decision rule is that the function $v$ defined by $v(t) = \E V_t$ is Bayes-plausible. Part \ref{it:initial_variance} is the \emph{initial variance constraint}: the sender's initial disclosure cannot increase the expected variance. Part \ref{it:dynamic_variance} is the \emph{no-disclosure upper bound}: the receiver's expected variance never increases faster than it would if the sender provided no information. For deterministic variance processes, which I call \emph{paths}, Bayes plausibility is also a sufficient condition, as I now show. 

To prove sufficiency,  I introduce \emph{delayed reporting}. At each time $t$, the sender reports the exact realization of the state at a previous time.  Initially, there is no previous state to report, so I define a fictitious history before time~$0$ to be used as a randomization device.  Let $Y = \{Y_t \}_{t \geq 0}$ be an independent standard Brownian motion. For $t < 0$, let $\th_t = \th_0 + Y_{-t}$. 
 
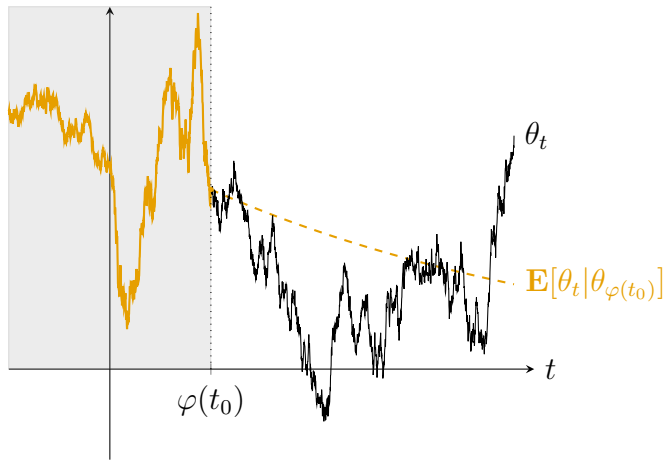
\begin{figure}
 \centering
 \def\t{0.5}
 \def\val{1.983}
 \begin{tikzpicture}
 \begin{axis}[
 	clip = false, 
 	scale only axis = true,
 	width = \textwidth/2,
 	axis x line = middle,
 	axis y line = middle,
 	xmin = -0.5,
 	xmax = 2.1,
 	ymin = -1,
 	ymax = 4,
 	xtick = {\t},
 	xticklabels = {$\varphi(t_0)$},
 	ymajorticks = false, 
 	every axis x label/.style={at={(current axis.right of origin)}, anchor=west},
 	xlabel = {$t$}
 ]

 	\filldraw [gray, opacity = 0.15] (-0.5,0) rectangle (\t, 4);
 
 	\addplot [thick, Orange, dashed, domain = \t:2, samples = 200] {\val*e^(-0.5*(x-\t))} node [pos = 1, anchor = west] {$\mathbf{E} [ \theta_t | \theta_{\varphi(t_0)}]$};
 	
 	\addplot [thick, Orange]  table [col sep = comma, x index = 0, y index = 1] {state_plot_a.csv};
 	\addplot [mark = none] table [col sep = comma, x index = 0, y index = 1] {state_plot_b.csv} node [pos = 1, anchor = west] {$\theta_t$};
 
 	\addplot [dotted] coordinates {(\t,0) (\t,4)};

 	\fill [Orange] (\t,\val) circle[radius=1pt, red];
 		
 \end{axis}
 \end{tikzpicture}	
 \caption{Delayed report at time $t_0$, and the receiver's updated expectations}
 \label{fig:delayed_reporting}
 \end{figure}

 Delayed reporting is parameterized by a \emph{reporting function}, defined to be a weakly increasing function $\varphi \colon [0, \infty) \to [-\infty, \infty)$ satisfying $\varphi(t) \leq t$ for each $t$.  At each time $t$, the sender reports $\th_{\varphi(t)}$. Let $\varphi(t) = -\infty$ if the sender has provided no information by time $t$. \cref{fig:delayed_reporting} shows a snapshot of the receiver's information under delayed reporting. At time $t_0$, the receiver's beliefs about $\th_t$ for $t \geq \varphi(t_0)$ depend only on the value of $\th_{\varphi(t_0)}$. The dashed curve shows the  conditional expectation function $\E [ \th_t | \th_{\varphi(t_0)}] = e^{\k (t - t_0)} \th_{\varphi(t_0)}$.

A reporting function $\varphi$ \emph{induces} a variance path $v$ if, for all $t \geq 0$,
\begin{equation*}
	 	\Var(\th_t | \th_{\varphi(t)}) = v(t),
\end{equation*} 
with the convention that $\Var(\th_t | \th_{-\infty}) = \Var(\th_t)$. 
 
 \begin{thm}[Delayed reporting]  \label{thm:delayed_reporting}
 Each Bayes-plausible variance path is induced by some reporting function. 
\end{thm}

If a variance path $v$ is induced by a reporting function $\varphi$, then for any bias path $b$, the following decision rule has deterministic bias $b$ and variance $v$: 
\begin{equation} \label{eq:bias_implementation}
	A_t = \E [ \th_t | \th_{\varphi(t)}] + b(t).
\end{equation}
Here, $A_t$ is random because the conditional expectation is random.

Delayed reporting simplifies the receiver's belief-updating process. Instead of aggregating the state information contained in the entire signal history, the receiver forms his belief at each time $t$ from the time-$t$ signal realization alone. Moreover, once $\varphi$ crosses $0$, the sender's signal is a deterministic function of the state history---the sender does not need to commit to randomization. One example of delayed reporting, with a different state process, is the optimal email notification policy in \cite{Ely2017}.\footnote{In \cite{Ely2017}, the state is binary, indicating whether an unread email is waiting. The optimal policy ``beeps'' after an email arrives, but with a delay of length $\D$. Thus, the reporting function is $\varphi(t) = \max \{ t - \D, 0\}$. In this setting, the initial state is known (no emails are waiting), so revealing $\th_0$ provides no information. Thus, there is no need for a fictitious history.}

\section{Optimal information policy} \label{sec:solution}

I now solve for the sender's optimal bias and variance functions. The associated decision rule with delayed reporting,  given in \eqref{eq:bias_implementation}, is a best response to the direct, grim-trigger information policy.

The sender's problem is to choose bias and variance functions $b$ and $v$ to solve
\begin{equation*}
\begin{aligned}
& \text{maximize} && -\int_{0}^{\infty} r e^{-r t} \big[ (b(t) - \b)^2 + v(t) \big] \de t \\
& \text{subject to} && 	-\int_{t}^{\infty} r e^{-r (s-t)} \big[ b^2(s) + v(s) \big] \de s \geq - \frac{\s^2 + r v(t)}{r - 2 \k}, \quad t \geq 0 \\
&&& 0 \leq v(s) \leq \h( v(t), s - t), \quad  s > t \geq 0\\
&&& 0\leq v(0) \leq \s_0^2.
\end{aligned}
\end{equation*}
The first constraint is obedience. The last two constraints impose Bayes plausibility. 

\begin{rem*}[Pareto frontier] Solving the sender's problem for arbitrary bias $\b$ immediately yields the entire Pareto frontier. Fix $\pi$ in $[0,1)$. Maximizing the social welfare function $\pi u_S + (1 - \pi) u_R$ is equivalent to maximizing the utility of a different sender with bias  $\pi \b$. This follows from the decomposition
\[
	\pi \brac{ ( b - \b)^2  +v } + (1 - \pi)  \brac{ b^2  + v } = (b - \pi \b)^2  + v + \pi (1 - \pi) \b^2. 
\]
\end{rem*}

\subsection{Obedience is binding}

Under the optimal policy, the obedience constraint must be active whenever the variance is strictly positive. Otherwise, the sender could strictly improve her payoff by reducing the variance $v$ over a small time interval where the constraint is slack. This perturbation relaxes the earlier obedience constraints and leaves the later obedience constraints unchanged.\footnote{In the formal proof, the optimal policy is derived directly, without first showing that obedience is active. The informal arguments in this section are intended to build intuition. }

The variance function, when positive (and differentiable), satisfies the differential equation 
\begin{equation} \label{eq:var_ODE}
	(r - 2 \k) b^2(t) = 2 \k v(t) + \s^2 -v'(t).
\end{equation}
This is the continuous-time analogue of the binding obedience constraint \eqref{eq:bias_variance_tradeoff} in the two-period example. On the right side, the first two terms capture the evolution of the receiver's posterior variance if he receives no additional information. The receiver is willing to bias his action by $b(t)$ only if the sender provides information that reduces his posterior variance (relative to its exogenous evolution) at rate $(r - 2 \k) b^2(t)$.  Hence, $r - 2 \k$ is the price (in  terms of variance) of inducing squared bias. This is the differential analogue of the price $\d^{-1} \rho^{-2}$ in the two-period example, with $\d = e^{-r}$ and $\rho = e^{\k}$. The higher the price, the more information the sender must initially withhold to induce a desired bias path. If the receiver is more impatient (higher $r$) or the process is less persistent (lower $\k$), the receiver demands a greater reduction in variance to select a given level of bias. 

If the variance first hits zero at some time $t$, then by continuity, the time-$t$ obedience constraint must be active. Starting at time $t$, the relaxed continuation problem, with subsequent obedience constraints dropped, is to choose $b(s)$ and $v(s)$ for $s > t$ to maximize the sender's continuation payoff subject to
\begin{equation*}
	\begin{aligned}
		-\int_t^{\infty} r e^{-r (s - t)} [b^2 (s)  + v(s)] \de s =  - \frac{\s^2}{r - 2 \k}.
	\end{aligned}
\end{equation*}
The sender's flow payoff is decreasing in $v$ and concave in $b$, so her continuation payoff is maximized by the stationary policy with $v (s) = 0$ and $b(s) = \min \{ \b, \s / \sqrt{r -2 \k}\}$, for $s > t$. This policy satisfies the obedience constraint at time $t$ and also at all subsequent times. Therefore, this policy is optimal in the original problem. It will form the stationary phase of the solution.

 \subsection{Optimal policy}

 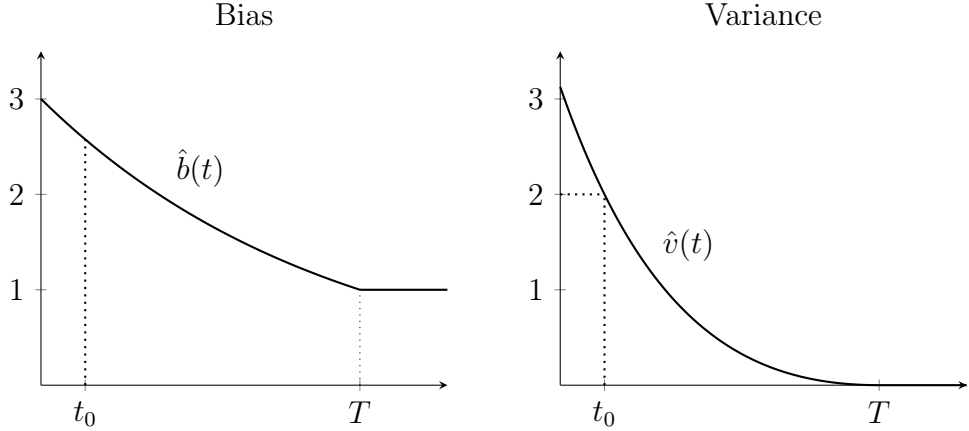
\begin{figure} 
 	\def\T{0.274653} 
 	\def\tf{0.35} 
 	\centering
 	\begin{tikzpicture}
 		\begin{groupplot}[
 			clip = false, 
 			axis x line = bottom,
 			axis y line = left,
 			xmin = 0,
 			xmax = \tf,
 			xtick = {0.0381103,\T},
 			xticklabels = {$t_0$,$T$},
 			ymin = 0,
 			ymax = 3.5,
 			ytick = {1,2,3},
 			yticklabels = {1,2,3},
 			group style = {group size = 2 by 1, horizontal sep = \plotgap},
 			]
 			
 			\nextgroupplot [title = Bias]
 			\addplot [thick, domain = 0:\tf, samples = 200] 
 			{3*(x < \T)*e^(-4*x) + (x >= \T)} node [pos = 0.5, above right] {$\hat{b}(t)$};
 			\addplot [thick, dotted] coordinates { (0.0381103, 0)  (0.0381103, 2.57583)};
 			
 			\addplot [dotted] coordinates {(\T,0) (\T,1)};
 			
 			\nextgroupplot [title = Variance]
 			\addplot [thick, domain = 0:\tf, samples = 200] 
 			{(x <= \T)*(4 + 2.28571*(-2*e^(\T -  x) + 0.25*e^(8*(\T - x))))} node [pos = 0.6, above right] {$\hat{v}(t)$};
 			\addplot [thick, dotted] coordinates { (0.0381103, 0)  (0.0381103, 2)};
 			\addplot [thick, dotted] coordinates { (0,2)  (0.0381103, 2)};

 		\end{groupplot}	
 	\end{tikzpicture}
 	\caption{Optimal bias and variance functions}
 	\label{fig:optimal_policy}
 \end{figure}

The main result characterizes the optimal bias and variance paths. 

\begin{thm}[Optimal bias and variance] \label{thm:optimal_policy}
The optimal bias $\hat{b}$ and variance $\hat{v}$ are unique and given as follows.\footnote{Here $x_+$ denotes the positive part $\max\{x,0\}$ of a real number $x$. In some expressions, the persistence parameter $\kappa$ appears in the denominator. The results still hold for $\k = 0$ if each expression is replaced with its limit as $\k$ tends to $0$.}
\begin{enumerate}[label = \Roman*.]
\item If $\b \leq \s / \sqrt{r - 2 \k}$, then $\hat{b}(t) = \b$ and $\hat{v}(t) = 0$.
\item If $\b > \s / \sqrt{r -2 \k}$, then 
\begin{align*} 
	\hat{b}(t)
	&= \frac{\s}{\sqrt{r - 2 \k}} e^{(r - 2 \k) (T-t)_+},\\
	\hat{v}(t) 
	&=- \frac{\s^2}{2 \k} + \frac{\s^2 (r - 2 \k)}{2(r - \k)} \Big( \k^{-1} e^{-2 \k (T - t)_+} + (r - 2 \k)^{-1} e^{2 ( r - 2\k) (T - t)_+} \Big),
\end{align*}
where the full-disclosure time $T$ takes the unique value for which the inequalities $\hat{b}(0) \leq \b$ and $\hat{v}(0) \leq \s_0^2$ both hold, at least one with equality.
\end{enumerate}
\end{thm}

There are two cases. If $\b \leq \s/ \sqrt{r - 2 \k}$, then the sender can induce her first-best decision rule $A_t = \th_t + \b$. At each time, the receiver's continuation value $-\b^2$ from this rule is weakly greater than his reservation utility $-\s^2 / (r -2 \k)$ from forfeiting all information and acting myopically. 

\begin{figure}
	\def\T{0.274653} 
	\def\tf{0.35} 
	\def\tint{0.133761} 
	\centering
	\begin{tikzpicture}
		\begin{axis}[
			clip = false, 
			xmin = 0,
			xmax = \tf,
			axis x line = bottom,
			axis y line = left,
			xtick = {\T},
			xticklabels = {$T$},
			ymin = 0,
			ymax = \tf,
			ytick = {1,2,3},
			yticklabels = {1,2,3},
			]
			
			\addplot [dashed] coordinates {(0,0) (\tf,\tf)} node [pos = 0.3, anchor = south east] {$t$};
			
			\addplot [thick, domain = 0.13:\tint, samples = 200] {(4*e^(x)- 1.56826*e^(8*x))/(e^(x) -1.1698435*e^(8*x))}; 
			\addplot [thick, domain = \tint:\T, samples = 200] {x + ln(-1.2857142858*e^(-8*x) + 1.5040845*e^(-x))} node [pos = 0.5, anchor = north west] {$\varphi(t)$}; 
			\addplot [thick] coordinates {(\tf, \tf) (\T,\T)}; 
			
			\addplot [dotted] coordinates {(\T,0) (\T,\T)};
			
		\end{axis}
	\end{tikzpicture}
	\caption{Optimal reporting function}
	\label{fig:reporting_function}
\end{figure}
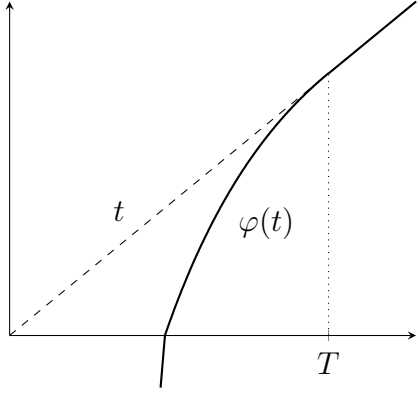

Hereafter, I focus on the interesting case in which $\b > \s / \sqrt{r -2 \k}$. The sender's first-best decision rule is not obedient, so the bias--precision tradeoff is in force. \cref{fig:optimal_policy} plots the optimal policy for a fixed set of parameters.\footnote{Here, $\b = 3$, $r = 3$, $\k = -0.5$, and $\s = 2$. The initial variance $\s_0^2$ is large enough that the initial variance constraint does not bind.} The optimal policy has two phases---a \emph{transition phase} until the full-disclosure time $T$, and a \emph{stationary phase} after time $T$. \cref{fig:reporting_function} plots the reporting function that implements this optimal policy. As time $t$ approaches $T$, the delay $t - \varphi(t)$ tends to $0$. Then $\varphi(t) = t$  for $t \geq T$. There is a kink when $\varphi$ crosses $0$ because the fictitious history has a different distribution than the true state process.

To derive the optimal policy, I first drop the initial variance constraint. I solve this relaxed problem by attaching a suitable multiplier to each time-$t$ obedience constraint. Then I integrate over these constraints to form the Lagrangian. In the solution of the relaxed problem, the optimal bias takes the form
\begin{equation} \label{eq:b_simple}
	\max \Set{ \b e^{-(r - 2\k) t}, \frac{\s}{\sqrt{r - 2 \k}} }.
\end{equation}
In \eqref{eq:b_simple}, the bias equals $\b$ at time $0$. At each time $t$ during the transition phase, the flow benefit of increasing the bias equals the shadow cost of reducing the variance. This shadow cost is initially zero. Over time, as the variance decreases,  this shadow cost increases, and the bias moves away from $\b$. The stationary phase begins at time $T$ when the variance hits zero and the bias hits $\s / \sqrt{r - 2 \k}$. 

If $\s_0^2$ is larger than the initial variance in the relaxed solution, then the relaxed solution solves the original problem. In this case, the sender makes an initial disclosure to reduce the receiver's variance below $\s_0^2$. Conversely, if $\s_0^2$ is smaller than the initial variance in the relaxed solution, then the relaxed solution is not feasible. Under the relaxed solution, the variance drifts downward and hits $\s_0^2$ at some time $t_0$. The solution of the original problem is the continuation policy from time $t_0$ onwards. In this case, there is no initial disclosure of information, and the sender never induces her preferred action bias $\b$. In \cref{fig:optimal_policy}, the time $t_0$ is indicated for initial variance $\s_0^2 = 2$. 

\begin{cor}[Optimum with deterministic state] \label{cor:optimal_deterministic}
If $\s = 0$, the optimal bias $\hat{b}$ and variance $\hat{v}$ are unique and given by
\begin{equation*} \label{eq:optimal_deterministic_bias_variance}
	\hat{b}(t) = b_0 e^{- (r - 2 \k )t},
	 \qquad \hat{v}(t) = \frac{r - 2 \k}{2(r - \k)} b_0^2 e^{-2 (r - 2 \k) t},
\end{equation*}
where $b_0 $ is the minimum of $\b$ and $\sqrt{ 2 \s_0^2 (r - \k)/(r - 2 \k)}$.
\end{cor}

This solution is the limit of the main solution as $\s$ tends to $0$. With $\s = 0$, the state evolves deterministically. If the receiver learns the current state, he can perfectly predict its future trajectory, so he will take his first-best decisions forever after.  When the action bias is zero, the receiver's marginal loss from increasing the action bias vanishes. Therefore, it is optimal for the sender to induce a slight bias by sacrificing some precision.\footnote{This contrasts with \citeapos{FudenbergRayo2019} transferable utility model. With $\s = 0$ and $\k = 0$, a variance path is Bayes-plausible if and only if it is weakly decreasing. This is identical to \citeapos{FudenbergRayo2019} constraint on the path of untransmitted knowledge. Under their optimal policy, however, this untransmitted knowledge hits zero in finite time. Thereafter, the agent keeps the whole surplus and chooses effort efficiently.}

\subsection{Comparative statics}

Say that a function \emph{increases} in response to a parameter change if it strictly increases at some point and does not decrease at any point.

\begin{prop}[Comparative statics] \label{prop:CS} 
Suppose that $\b > \s / \sqrt{r - 2 \k}$ and the initial variance constraint is not active.
\begin{enumerate}
	\item \label{it:cs_bias} The optimal bias function $\hat{b}$ is increasing in $\b$ and $\s$ and decreasing in $r - 2 \k$.
	\item \label{it:cs_variance} The optimal variance function $\hat{v}$ is increasing in $\b$ and decreasing in $\s$. 
\end{enumerate}
\end{prop}

In the stationary phase, $\hat{b}(t) = \s / \sqrt{r - 2 \k}$ and $\hat{v}(t) = 0$, so the bias is increasing in $\s$ and decreasing in $r - 2\k$. In the transition phase, the comparative statics for the bias and variance paths are the same as those for the first-period bias and variance in the two-period example (with discount factor $\d = e^{-r }$ and discrete-time persistence $\rho = e^{\k}$). As the preference bias $\b$ increases, the sender withholds more information to induce greater action bias. As the volatility $\s$ increases, the \emph{sensitivity} of the receiver's reservation utility to the variance does not change, but the \emph{level} of the receiver's reservation utility decreases, relaxing the obedience constraint. Thus, the optimal bias $\hat{b}$ over the transition phase does not change, but the variance $\hat{v}$ decreases.\footnote{The shadow cost of reducing variance remains the same because of the linearity assumptions: payoffs are linear in the variance, and the state follows a linear stochastic differential equation. If the initial variance constraint binds, then $\hat{b}$ does increase in $\s$ over the transition phase.} Recall from \eqref{eq:var_ODE} that $r -2 \k$ is the price (in variance) of inducing bias. As this price increases, the bias decreases. The effect of $r$ and $\k$ on the variance, however, is ambiguous. As $r$ increases and $\k$ decreases, more variance is required to induce a given level of bias, but the optimal bias path is lower. Which effect  dominates depends on other parameter values and can change over time.

\section{Multidimensional states and actions} \label{sec:multidim}

The main model studies \emph{how much} information the sender provides over time. I now consider a multidimensional state in order to study \emph{which} information the sender provides at each time. Returning to one of the motivating examples, suppose that the main unit in an organization has private information about two evolving situations that are relevant to another unit. If the main unit wants to extract concessions from the other unit, the solution below suggests that it is optimal to withhold information about the less mean-reverting situation for longer, revealing it only after fully disclosing the more mean-reverting situation. 

Suppose that the state is $n$-dimensional, denoted $\th_t = (\th_{t,1}, \ldots, \th_{t,n}) \in \R^n$. The components of the initial state $\th_0$ are independent. Each component $\th_{0,i}$ has a normal distribution $N(\mu_i, \s_{0,i}^2)$. Thereafter, the components evolve independently. Each component $\th_{t,i}$ follows the linear stochastic differential equation
\[
	\de \th_{t,i} = \k_i \th_{t,i} \de t + \s_{i} \de Z_{t,i},
\]
where $Z_i = \{ Z_{t,i}\}_{t \geq 0}$ is a standard Brownian motion,  and $Z_1, \ldots, Z_n$ and $\th_0$ are mutually independent. For each $i$, assume $\s_i > 0$ and $2 \k_i < r$.  Order the components by increasing persistence, so $\k_1 \leq \cdots \leq \k_n$.

At each time $t$, the receiver chooses an action $a_t = (a_{t,1}, \ldots, a_{t,n}) \in \R^n$. The flow utilities for the sender and receiver are given by
\begin{equation*}
	u_S (a_t, \th_t) = - \sum_{i=1}^{n} ( a_{t,i} - \th_{t,i} - \b_i)^2,
	\qquad
	u_R (a_t, \th_t) = - \sum_{i=1}^{n} (a_{t,i} - \th_{t,i})^2.
\end{equation*}
The sender's preference bias is a vector $\b = (\b_1, \ldots, \b_n) \in \R^n$. 

With minor modifications to the argument in the main model, it can be shown that the optimal policy induces deterministic bias and variance functions, which are now vector-valued. For each $i$, let
\[
	b_i(t) = A_{t,i} - \E [ \th_{t,i} | \FF_t^A]
	\quad
	\text{and}
	\quad
	v_i(t) = \var (\th_{t,i} | \FF_t^A).
\]
Write $b(t) = (b_1(t), \ldots, b_n(t))$ and $v(t) = (v_1(t), \ldots, v_n(t))$. Define a separate variance-updating function $\h_i $ as in \eqref{eq:var_exp} for each component $i$. 

The sender chooses vector-valued bias and variance functions $b$ and $v$ to solve
\begin{equation*} \label{eq:multidim_sender_problem}
\begin{aligned}
	& \text{maximize} 
	&&  -\int_{0}^{\infty} r e^{-r t} \sum_{i=1}^{n} \big[ (b_i(t) - \b_i)^2 + v_i(t) \big]\de t \\
	&\text{subject to} 
	&& -\int_{t}^{\infty} r e^{-r(s - t)} \sum_{i=1}^{n}  \big[ b_i^2(s) +v_i(s) \big] \de s \geq -\sum_{i=1}^{n} \frac{ \s_i^2 + r v_i(t)}{r - 2 \k_i}, \quad t \geq 0 \\
	&&& 0 \leq v_i(s) \leq \h_i ( v_i(t), s - t), \quad i = 1, \ldots, n, \quad s > t \geq 0 \\
	&&&0\leq v_i(0) \leq \s_{0,i}^2, \quad i = 1, \ldots, n.
\end{aligned}
\end{equation*} 
The time-$t$ obedience constraint is the sum of the single-dimensional time-$t$ obedience constraints for each component. It is feasible to separately choose the single-dimensional optimal policy for each component, but this is generally suboptimal.

As in the main model, the obedience constraint must be active at each time $t$. The variance vector (when differentiable) satisfies the differential equation 
\begin{equation} \label{eq:multi-dim_obedience}
	\| b (t) \|^2  = \sum_{i=1}^{n} \frac{2 \k_i v_i(t) + \s_i^2 - v_i'(t)}{r - 2 \k_i}.
\end{equation}
This is the multidimensional analogue of \eqref{eq:var_ODE}. Only the magnitude of the bias vector appears in the obedience constraint, so it is optimal to always choose $b(t)$ parallel to $\b$. With a single-dimensional state, bias has price $r - 2 \k$ in units of variance, which is the only currency. With $n$ state components, there are $n$ currencies. The price of bias is $r -2 \k_i$ in units of $v_i$. The volatility parameters do not enter the price because they do not affect the \emph{sensitivity} of the reservation utility to the current variance. Since the variances $v_1, \ldots, v_n$ have the same effect on the sender's flow payoff, the sender prefers to pay in the currency with the lowest price, i.e., the highest $\k_i$. If the initial variance $\s_{0,n}^2$ is sufficiently large, the sender compensates the agent with information about only the most persistent component $n$. All other components are revealed immediately. 
If $\s_{0,n}^2$ is too small, then the sender must withhold information about other components as well. The components are revealed sequentially, in order of increasing persistence, so that the largest bill is paid at the lowest price. 

To state the theorem,  let $\hat{\s}_j^2 = \sum_{i=1}^{j} \s_i^2/(r - 2 \k_i )$ for $j = 1, \ldots, n$. Set $\hat{\s}  = \hat{\s}_n$. Observe from \eqref{eq:multi-dim_obedience} that $\hat{\s}$ is the magnitude of bias that the sender can induce while keeping the receiver perfectly informed. 

\begin{thm}[Optimum with multidimensional state] \label{thm:multidim} 
If the state is $n$-dimensional, then the following vector-valued bias and variance paths, $\hat{b}$ and $\hat{v}$, are optimal. The optimum is unique if $\k_1 <  \cdots < \k_n$.
\begin{enumerate}[label = \Roman*.]
\item If $\| \b\| \leq \hat{\s}$, then $\hat{b}(t) = \b$ and $\hat{v} (t) = 0$. 
\item If $\| \b\| > \hat{\s}$, then for some uniquely determined critical component $i_0$ and full-disclosure times $0 = t_1 = \cdots =  t_{i_0 - 1} < t_{i_0}  \cdots < t_n$,\footnote{If $\s_{0,i}^2=  0$ for all $i$, then $(t_1, \ldots, t_n) = 0$, so technically $i_0 = n + 1$. Outside of this edge case, we have $i_0 \leq n$ and the full-disclosure times are pinned down by the following conditions:
\begin{enumerate}[label = (\roman*), ref = \roman*]
	\item $\| \hat{b}(0)  \| \leq \|\b\|$, $\hat{v}_{i_0}(0) \leq \s_{0,i_0}^2$, and $\hat{v}_i(0) = \s_{0,i}^2$ for $i > i_0$;
	\item \label{it:multidim_equal} $\| \hat{b}(0) \| = \| \b\| $ or $(i_0, \hat{v}_1(0)) = (1, \s_{0,1}^2)$.
\end{enumerate}}
\[
	\hat{b} (t) = \hat{\s} \exp \Paren{ \sum\nolimits_{i=i_0}^{n} (r - 2 \k_i) (t_i - t \vee t_{i-1})_+}  \frac{\b}{ \| \b \|}.
\]
For $i < i_0$, we have $\hat{v}_i(t) = 0$. For $i  \geq i_0$, the variance $\hat{v}_i$ is defined piecewise. For $t \geq t_{i-1}$,
\begin{multline} \label{eq:hatv_multi}
	\hat{v}_i(t) 
	= \frac{\hat{\s}_{i}^2 (r - 2 \k_i) }{2 \k_i} \Big( e^{-2 \k_i (t_i - t)_+} - 1 \Big) \\
	+ \frac{\hat{\s}^2 (r - 2 \k_i) e^{2 \tau_{i+1}}}{2 ( r- \k_i)}  \Big( e^{2( r - 2\k_i) (t_i - t)_+} - e^{-2 \k_i (t_i - t)_+} \Big),
\end{multline}
where $\tau_{j} = \sum_{i=j}^{n} (r - 2 \k_i) (t_i - t_{i-1})$ for $j = 1 \ldots, n$.  For $t < t_{i-1}$, 
\[
	\hat{v}_i( t) =  - \frac{\s_i^2}{2 \k_i} + \Paren{\hat{v}_i(t_{i-1}) + \frac{\s_i^2}{2 \k_i}} e^{2 \k_i (t - t_{i-1})}.
\]
\end{enumerate}
\end{thm}

\begin{figure} 
	\def\tii{0.0418251}
	\def\tiii{0.363075}
	\def\tf{0.55}	
	\def\biii{1.93465}
	\def\bii{4.31911}
	\def\vii{4.25359}
	\centering	
	
	\begin{tikzpicture}
		\begin{groupplot}[
			clip = false, 
			axis x line = bottom,
			axis y line = left,
			xmin = 0,
			xtick = {\tii,\tiii},
			xticklabels = {$t_2$,$t_3$},
			xmax = \tf,
			ymin = 0,
			ytick = {1, 2, 3,4, 5},
			yticklabels = {1,2,3,4,5},
			ymax = 5.5,
			group style = {group size = 2 by 1, horizontal sep = \plotgap},
			]
			\nextgroupplot[title = Bias magnitude]
			\addplot[thick, domain = 0:\tii]{5*e^(-3.5*x)};
			\addplot[thick, domain = \tii:\tiii]{4.79519*e^(-2.5*x)};
			\addplot[thick, domain = \tiii:\tf] {1.93465} node [anchor = south, pos = 0.5] {$\|\hat{b}(t)\|$};
			
			\addplot[dotted] coordinates {(\tii,0) (\tii,\bii)};
			\addplot[dotted] coordinates {(\tiii,0) (\tiii,\biii)};
			
			\nextgroupplot[title = Variances]
			\addplot[thick, Orange, domain = 0:\tii] {15 + 13.4615*e^(-7*x) - 25.5741*e^(-x/2))};
			\addplot[thick, Orange, domain = \tii:\tf] {0};
			
			\addplot[thick, Blue, domain = 0:\tii] {-8. + 12*e^(x/2)};
			\addplot[thick, Blue, domain = \tii:\tiii] {-18.7143 + 10.4517*e^(-5*x) + 14.1886* e^(x/2)};
			\addplot[thick, Blue, domain = \tiii:\tf] {0};
			
			\addplot [dotted] coordinates {(\tii,0) (\tii,\vii)};
			
			\draw (0.12, 0) node [anchor = south] {\textcolor{Orange}{$\hat{v}_2(t)$}};
			\draw (0.12,2.9)node [anchor = south] {\textcolor{Blue}{$\hat{v}_3(t)$}};
		\end{groupplot}	
	\end{tikzpicture}
	\caption{Optimal information policy with three-dimensional state}
	\label{fig:multidim_sol}
\end{figure}
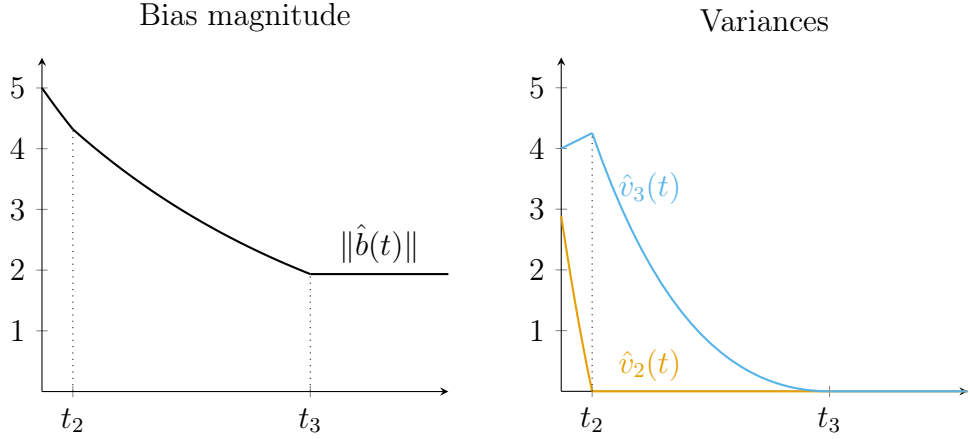

I focus on the interesting case in which $\| \b\|  > \hat{\s}$. Figure \ref{fig:multidim_sol} plots the optimal policy in an example with three components.\footnote{Here, $\| \b\|  = 5$, $r = 3$, $(\k_1, \k_2, \k_3) = (-0.75, -0.25, 0.25)$, and $\s_1 = \s_2 = \s_3 =  \s_{0,3} = 2$.  The initial variance $\s_{0,2}^2$ is large enough that the initial variance constraint for state $2$ does not bind. The value of $\s_{0,3}^2$ does not affect the solution.} (The variance $v_1(t)$ is always zero, so it is not plotted.) The critical component is $i_0 = 2$. At time $t_1= 0$, the sender fully discloses component $1$ and makes a partial disclosure about component $2$. Then the sender gradually provides information about component $2$, while keeping the receiver fully informed of component $1$ and providing no information about component $3$.  At time $t_2$, component $2$ is fully revealed. Then the sender gradually reveals component $3$, while keeping the receiver fully informed of components $1$ and $2$. At time $t_3$, the entire state is fully revealed, and the stationary phase begins: $\| \hat{b}(t) \| = \hat{\s}$ and $\hat{v}(t) = 0$ for $t \geq t_3$. Between the threshold times $t_{i-1}$ and $t_i$, the bias magnitude decays exponentially at rate $r - 2 \k_i$, the optimal rate of decay in the single-dimensional model with component $i$ as the state.

\section{Conclusion} \label{sec:conclusion}

This paper studies the optimal provision of information in a long-term relationship. The sender cannot induce actions that are biased in her own favor and also precisely tailored to the state. She resolves this tradeoff with a non-stationary policy that reveals information gradually over time. Initially, actions are biased but imprecise. Over a transition phase, actions become more precise but less biased, eventually reaching a stationary phase of perfect precision and constant bias. Throughout the relationship, the receiver chooses these biased actions in order to continue receiving information from the sender. The dynamics of this informational relationship arise endogenously, not because the sender comes to “trust” the receiver, but as the solution to a dynamic incentive problem.

\appendix

\newpage

\section{Formal definition of strategies} \label{sec:strategies}

Fix a probability space $(\W, \FF, \P)$ that is sufficiently rich to carry all the random objects introduced below. The driving process $Z = \{ Z_t \}_{t \geq 0}$ is a standard Brownian motion. The initial state $\th_0$ has a normal distribution $N(\mu_0, \s_0^2)$, independent of $Z$. Let
\begin{equation} \label{eq:stochastic_integral}
	\th_t = \th_0 e^{\k t} + \s \int_{0}^{t} e^{\k (t - s) } \de Z_s,
\end{equation}
where the integral is an It\^{o} integral. This process satisfies the stochastic differential equation in \eqref{eq:SDE}. Let $\{ \FF_t^\th \}$ denote the filtration generated by $\{\th_t \}$,%
\footnote{That is, $\FF_t^\th = \s ( \th_s: s \leq t)$ for each $t$.}
and set $\FF_{\infty}^{\th} = \s ( \bigcup_{t \geq 0} \FF_t^\th )$.

As a randomization device, the sender is endowed with a standard Brownian motion $Y = \{Y_t\}_{t \geq 0}$ that is independent of $\FF_\infty^\th$.  Let $\tilde{\FF}_t^\th = \s ( \s( Y) \cup \FF_t^\th)$. The filtration $\{ \tilde{\FF}_t^\th \}$ represents the exogenous information available to the sender. 

To define the receiver's action paths, let  $\operatorname{RC}[0,\infty)$ denote the space of all right-continuous functions from $[0,\infty)$ to $\R$. Equip $\operatorname{RC}[0, \infty)$ with the cylindrical filtration $\{ \CC_{t-} \}$, where $\CC_{t-}$ denotes the $\s$-algebra generated by the projection maps for times strictly before $t$. 

An \textit{information policy} consists of a measurable space $(\mathbf{S}, \SS)$ and a map
\[
	S \colon [0, \infty) \times \W \times \operatorname{RC}[0, \infty) \to \mathbf{S}
\]
that is adapted to the filtration $ \{ \tilde{\FF}_t^\th \otimes \CC_{t-} \}$ on $\W \times \operatorname{RC}[0, \infty)$. The interpretation is that the sender's signal at time $t$ can depend on (i) the sender's exogenous information, through $\tilde{\FF}_t^\th$, and (ii) actions taken by the receiver strictly before time $t$, through $\CC_{t-}$. Denote the entire information policy by $S$.

A \textit{decision rule} is a real-valued $\{ \tilde{\FF}_t^\th \}$-adapted stochastic process on $(\W, \FF)$ with right-continuous sample paths. By \citet[][Proposition 1.13, p.~5]{KaratzasShreve1998}, a decision rule is progressively measurable with respect to $\{ \tilde{\FF}_t^\th \}$.

Next, I define the compatibility of a decision rule with an information policy. Once this definition is in place, say that a decision rule $A$ is a \emph{best response} to an information policy $S$ if (i) $A$ is compatible with $S$, and (ii) $u_R(A) \geq u_R(A')$ for all decision rules $A'$ compatible with $S$.  Unfortunately, perfect monitoring in continuous time poses technical challenges \citep{SimonStinchcombe1989,BerginMacLeod1993}. In settings with continuous actions, I am not aware of a satisfactory restriction on strategies that avoids all of these technical problems.\footnote{The challenge is that the sender chooses how her signals depend on the receiver's actions in the \emph{arbitrarily recent} past. If the dependence of the signals on actions is exogenous and noisy, e.g., through the drift of a Brownian motion as in \citet{Sannikov2008}, then decision rules can be defined with respect to an exogenous filtration. Without this exogenous structure, the measurability conditions become self-referential. Without further restrictions, these measurability conditions do not exclude pathological decision rules in which the receiver uses the sender's feedback rule to instantaneously transmit to himself exogenous information that he never receives directly. 
	
Existing methods cover only settings in which the players choose when to switch between discrete actions. The grid method of \cite{SimonStinchcombe1989} and the inertia strategies of \cite{BerginMacLeod1993} are defined in deterministic environments.  \cite{KamadaRao2021} introduce a new approach for stochastic settings.} Fortunately, the obedience characterization  (\cref{res:obedience}) is robust to the exact definition of strategies. To demonstrate this, I take the following axiomatic approach.

 The sender is restricted to some subcollection of \emph{admissible} information policies. Each admissible information policy is associated with a collection of decision rules that are \emph{compatible} with that policy. I assume that these notions of admissibility and compatibility satisfy the following conditions. 
\begin{enumerate}[ label = C\arabic*., ref = C\arabic* ]
	\item \label{it:modification} If a decision rule $A$ is compatible with an admissible information policy $S$, then so is any decision rule $A'$ defined as follows. For some fixed time $t_0$ and some event $G$ in $\FF_{t_0}^A$,  let $A_t'(\w) = A_t(\w)$ whenever $t < t_0$ or $\w \notin G$; otherwise, define $A'$ so that $G A_t'$ is $\FF_{t_0}^A$-measurable for all $t \geq t_0$.\footnote{Here and below, I denote a set and its indicator function by the same symbol.}
	
	\item  \label{it:trigger_feasible} The following trigger information policies are all admissible. Let $X$ be a decision rule. Let $T \colon \operatorname{RC} [0, \infty) \times \operatorname{RC} [0, \infty) \to \R$ be a function such that, for each time $t$, the event $[T < t]$ is in $\CC_{t-} \otimes \CC_{t-} $. The $(X,T)$-trigger information policy $S$ is defined by 
	\[
		S(t, \w, a) = X_{ t \wedge T ( X(\w), a)} (\w). 
	\]
	Since $X$ is $\{ \tilde{\FF}_{t}^{\th} \}$-progressively measurable, $S$ is $\{ \tilde{\FF}_{t}^\th \otimes \CC_{t-}\}$-adapted. 
	
	\item \label{it:trigger} A decision rule $A$ is compatible with the $(X,T)$-trigger policy $S$ if and only if
	\[
		[ A_t \in B] \cap [ T(X, A) \wedge t < s] \in \FF_s^X,
	\]
	for every Borel set $B$ and all times $s$ and $t$. This implies, in particular, that $A$ is $\{ \FF_t^X\}$-adapted.

\end{enumerate}

Condition \ref{it:modification} means that if the receiver can follow the decision rule $A$ under information policy $S$, then starting at time $t_0$, conditional on the event $G$, the receiver can select different actions, using information that was available at time $t_0$.
Condition \ref{it:trigger} means that the receiver's decision cannot depend on the realizations of the process $X$ after it is stopped by the sender. 

It can be checked that the definition  of compatibility (with trigger policies) in \ref{it:trigger} satisfies \ref{it:modification}. Thus, it is consistent with the axioms to restrict the sender to trigger information policies and to define compatibility by \ref{it:trigger}. Of course, there are many other classes of reasonable information policies. If we include additional policies in the admissible set, then the obedience characterization (\cref{res:obedience}) still holds, as long as compatibility with these new information policies is defined in a way that is consistent with \ref{it:modification}.

\section{Proofs} \label{sec:proofs}

\subsection{Preliminaries}

In the proofs below, I use the following form of the law of total variance. For any square-integrable random variable $X$ and any sub-$\s$-algebras $\GG$ and $\HH$ satisfying $\GG \supset \HH$, 
\begin{equation} \label{eq:strong_total_var}
	\Var ( X | \HH) = \E [ \var(X| \GG) | \HH] + \var ( \E [ X | \GG] | \HH ) \geq \E [ \var (X | \GG)  | \HH].
\end{equation}
Taking $\HH$ to be the trivial $\s$-algebra gives the usual law of total variance. 

If $X$ and $Y$ are square integrable, and $X$ is measurable with respect to $\GG$, then
\begin{equation} \label{eq:quadratic_comparison}
	\E (X - Y)^2 \geq \var ( X - Y) \geq \E [ \Var  (X - Y| \GG)] = \E [ \var ( Y | \GG)],
\end{equation}
where the middle inequality uses the usual law of total variance and the last inequality uses the $\GG$-measurabilty of $X$. 

\subsection{Proof of \texorpdfstring{\cref{res:obedience}}{Proposition \ref{res:obedience}}}

First, I prove that obedience is necessary. I prove the contrapositive. Let $A$ be a decision rule that is not obedient. That is, there exists some time $t$ and some positive-measure set $G$ in $\FF_t^A$ such that
\begin{equation} \label{eq:obedience_violation}
	- \E \Brac{ G \int_{t}^{\infty} r e^{-r (s - t)} (A_s - \th_s)^2 \de s} < -\E \Brac{ G  \frac{ \s^2 + r (\E [ \th_t | \FF_t^A] - \th_t )^2}{r -2 \k} }.
\end{equation} 
Define a new decision rule $A'$ by setting $A_s' = e^{ \k (s - t)} \E [ \th_t | \FF_t^A]$ for $s \geq t$ on $G$, and setting $A'$ equal to $A$ otherwise. By \eqref{eq:obedience_violation}, it follows that $u_R (A') > u_R (A)$. Whenever $A$ is compatible with an information policy, then so is $A'$ (by \ref{it:modification}), so $A$ cannot be a best response to any policy. 

In order to prove that obedience is sufficient, I first define the direct, grim-trigger information policy $S$ associated to a fixed decision rule $A$ as the trigger decision rule (from \ref{it:trigger_feasible}) with $X = A$ and 
\[
	T(a, a') =  \inf \Set{ t \geq 0: \int_{0}^{t} (a_s - a_s')^2 \de s > 0 },
\]
where the infimum of the empty set equals $\infty$.  Using Fubini's theorem, it can be shown that the event $[T < t]$ is in $\{ \CC_{t-} \otimes \CC_{t-} \}$, as required. 
The information policy $S$ is defined by 
\[
	S(t, \w, a) = A_{t \wedge T ( A(\w), a)} (\w). 
\]

 Now I prove that obedience is sufficient. Let $A$ be an obedient decision rule. Let $S$ be the associated direct, grim-trigger information policy. I claim that $A$ is a best response to $S$. Clearly $A$ is compatible with $S$, since $T( A(\w), A(\w)) = \infty$ for all $\w$. Let $A' = \{ A'_t\}$ be an arbitrary decision rule that is compatible with $S$.  I claim that $u_R (A) \geq u_R (A')$. This inequality holds trivially if $u_R (A') = -\infty$, so assume that $u_R (A')$ is finite.
 
Define the random time $T'$ by $T'(\w) = T(A(\w), A'(\w))$. I approximate $T'$ from above by simple functions. For each $n$, define the simple random times $T_n$ by 
\[
T_n
= \begin{cases}
	j / 2^{n} &\text{if}~ (j-1) /2^{n} \leq T' <  j/2^{n}, \; j = 1, \ldots, n 2^n,\\
	\infty &\text{if}~ T' \geq n.
\end{cases}
\]
By construction,  $T_n > T'$ and $T_n \downarrow T'$. Let $A^n$ be the decision rule that agrees with $A$ if $t < T_n$ and agrees with $A'$ if $t \geq T_n$. Since $u_R(A')$ and $u_R(A)$ are both finite, Lebesgue's dominated convergence theorem implies that $u_R(A^n) \to u_R(A')$. Therefore, it suffices to check that $u_R(A^n) \leq u_R(A)$ for each $n$. 

Fix $n$ and let $t_1, \ldots, t_K$ denote the finite values that $T_n$ takes with positive probability. For each $k$, let $I_k$ be the indicator for the event that $T_n = t_k$. We have
\begin{equation} \label{eq:sum_exp}
	u_R (A) - u_R(A^n) =  \sum_{k=1}^{K} \E \Brac{ I_k \int_{t_k}^{\infty} r e^{-r t} \Paren{ ( A_t' - \th_t)^2  -  (A_t - \th_t)^2} \de t }. 
\end{equation}
For $t \geq t_k$, the random variable $I_k A_t'$ is $\FF_{t_k}^A$-measurable by \ref{it:trigger}, so \eqref{eq:quadratic_comparison} gives
\begin{equation} \label{eq:key_ineq}
	\E [ I_k (A_t' - \th_t)^2] \geq \E [ I_k \Var ( \th_t | \FF_{t_k})] = \E[ I_k \h ( \Var ( \th_{t_k} | \FF_{t_k}), t - t_k)],
\end{equation}
where $\h$ is the function defined in \eqref{eq:var_exp}. To see that each expectation in \eqref{eq:sum_exp} is nonnegative, change the order of integration, substitute in \eqref{eq:key_ineq}, and then use the time-$t_k$ obedience constraint for $A$. We conclude that $u_R (A) \geq u_R(A^n)$. 

\subsection{Proof of \texorpdfstring{\cref{res:bias_variance}}{Proposition \ref{res:bias_variance}}} \label{sec:proof_bias_variance}
Let $A$ be an obedient decision rule. Define the bias and variance functions $b$ and $v$ by
\[
	b(t) = \E [ A_t - \th_t] 
	\quad \text{and} \quad
	v(t) = \E [ \Var(\th_t | \FF_t^A)].
\]
First, I check that $v$ is Bayes-plausible. By the law of total variance, 
\[
	v(0) = \E [  \var (\th_0 | \FF_0^A) ] \leq \var( \th_0) = \s_0^2. 
\]
For $s > t$, the stronger law of total variance in \eqref{eq:strong_total_var} gives
\begin{equation} \label{eq:st_var}
	\E [ \var ( \th_{s} | \FF_s^A ) | \FF_t^A] \leq \var( \th_{s} | \FF_t^A) = \h  (\var( \th_t | \FF_t^A), s - t),
\end{equation}
where the equality can be derived from \eqref{eq:stochastic_integral}, using 
It\^{o}'s isometry and the $\{ \tilde{\FF}_t^{\th}\}$-adaptedness of $A$. Since $\h$ is linear in its first argument, taking expectations in \eqref{eq:st_var} gives $v(s) \leq \h (v(t), s - t)$.

Since $v$ is Bayes-plausible, it follows from \cref{thm:delayed_reporting} (proven below, without appealing to \cref{res:bias_variance}) that there exists a reporting function $\varphi$ that induces $v$. Define the decision rule $A'$ by 
\[
	A_t' = \E [ \th_t | \th_{\varphi(t)}] + b(t). 
\]
Take expectations in the time-$t$ obedience constraint for $A$ to get
\[	
	\E \Brac{ \int_{t}^{\infty} r e^{-r (s - t)} (A_s' - \th_s)^2 \de s} \leq \frac{ \s^2 + r \E [\Var(\th_t | \th_{\varphi(t)})]}{r - 2 \k},
\]
which is exactly the time-$t$ obedience constraint for $A'$. Thus, $A'$ is obedient.

It remains to check that both players weakly prefer $A'$ to $A$. This holds because
\[
	\E ( A_t - \th_t)^2  
	= \E[ B_t^2]  + \E[ V_t] 
	 \geq (\E B_t)^2 + \E[V_t] 
	 = \E (A_t'- \th_t)^2,
\]
and similarly,  $\E ( A_t - \th_t - \b)^2 \geq  \E (A_t'- \th_t - \b)^2$.

\subsection{Proof of \texorpdfstring{\cref{thm:delayed_reporting}}{Theorem \ref{thm:delayed_reporting}}} \label{subsec:proofs_delayed_reporting}

Let $v \colon [0, \infty) \to [0, \infty)$ be Bayes-plausible. Then $v$ has the following \emph{monotonicity} property. If $v(t) \leq \h(w,t)$ for some fixed $w$ and $t$, then for $s > t$ we have
\begin{equation*}
	v(s) \leq \h(v(t), s - t) \leq \h( \h(w,t), s- t ) = \h( w, s),
\end{equation*}
where the second inequality holds because $\h$ is strictly increasing in its first argument. Moreover, the second inequality holds strictly if $v(t) < \h (w, t)$. 

Define the reporting function $\varphi$ implicitly by the following piecewise system (which separates into cases according to the sign of $\varphi(t)$):
\[
\begin{cases} 
\varphi(t)  = - \infty &\text{if}~ v(t) = \h(\s_0^2, t),\\
v(t) = \h\big( (1/\s_0^2 - 1/\varphi(t))^{-1}, t \big) 	&\text{if}~ \h(0,t) < v(t) < \h(\s_0^2, t),\\
v(t) = \h(0, t - \varphi(t))	 &\text{if}~0 \leq v(t) \leq \h(0,t).
\end{cases}
\]
This construction ensures that $\varphi$ induces $v$, provided that $\varphi$ is a well-defined reporting function. 

First I check that $\varphi$ is well-defined. By the initial variance constraint and the monotonicity property, $v(t) \leq \h (\s_0^2, t)$ for all $t \geq 0$, so the cases are exhaustive. In the second case, the solution  is unique because $\h$ is strictly increasing in its first argument. In the third case, the solution is unique because $\h(0, \cdot)$ is strictly increasing and $\h(0, 0) = 0$. 

Now I check that $\varphi$ is a reporting function. For a fixed $t$, if the second case obtains, we get $\varphi(t) \leq 0$. If the third case obtains, we get $0 \leq \varphi(t) \leq t$.  The monotonicity property ensures that as time passes, it is only possible to move from an earlier case to a later case (and not the reverse). Within each case, the monotonicity property ensures that $v$ is weakly increasing in time (since $\h$ is strictly increasing in its first argument and the function $\h(0, \cdot)$ is strictly increasing). 

\subsection{Proof of  \texorpdfstring{\cref{thm:optimal_policy}}{Theorem \ref{thm:optimal_policy}}} \label{subsec:proof_optimal_policy}

Assume $\b > \s / \sqrt{r - 2 \k}$, for otherwise the result is clear. The full-disclosure time $T$ in the theorem statement is well-defined because the expressions for $\hat{v}(0)$ and $\hat{b}(0)$, as functions of $T$, are strictly increasing.

 Drop the no-disclosure bounds on the variance to obtain the relaxed problem\footnote{To simplify notation, I work with losses rather than utilities throughout the appendix.}
\begin{equation} \label{eq:proof_optimal_policy_relaxed}
\begin{aligned}
&\text{minimize} 
&& \int_{0}^{\infty} r e^{-r t} \big[ (b(t) - \b)^2 + v(t) \big] \de t \\
&\text{subject to} 
&& \int_{t}^{\infty} r e^{-r (s - t)} \big[ b(s)^2 + v(s) \big] \de s
\leq \frac{ \s^2 + r v(t)}{r - 2 \k}, \quad t \geq 0 \\
&&& v(t) \geq 0, \quad t \geq 0 \\
&&& v(0) \leq \s_0^2.
\end{aligned}
\end{equation}
It is straightforward to check that $\hat{v}$ satisfies the no-disclosure bounds. I will prove that $(\hat{b}, \hat{v})$ is the unique solution of \eqref{eq:proof_optimal_policy_relaxed}. The proof is separated into two parts. The first part proves optimality. In the second part, the claimed uniqueness is stated precisely and then proved. 

\paragraph{Optimality} 
First, drop the initial variance constraint:
\begin{equation} \label{eq:proof_optimal_policy_aux}
\begin{aligned}
&\text{minimize} 
&& \int_{t_0}^{\infty} r e^{-r t} \big[ (b(t) - \b)^2 + v(t) \big] \de t \\
&\text{subject to} 
&& \int_{t}^{\infty} r e^{-r (s - t)} \big[ b^2(s) + v(s) \big] \de s
\leq \frac{ \s^2 + r v(t)}{r - 2 \k}, \quad t \geq 0 \\
&&& v(t) \geq 0, \quad t \geq 0. 
\end{aligned}
\end{equation}

 Define auxiliary functions $\tilde{b}$ and $\tilde{v}$ by the corresponding expressions for $b$ and $v$ in the theorem statement, but with $T$ defined by
 \begin{equation} \label{eq:modified_T}
 	\b e^{-(r - 2\k) T} = \frac{\s}{\sqrt{r - 2 \k}}.
 \end{equation}
Below, I will show that $(\tilde{b}, \tilde{v})$ solves \eqref{eq:proof_optimal_policy_aux}.  I claim that this implies that $(\hat{b}, \hat{v})$ solves \eqref{eq:proof_optimal_policy_relaxed}. If $\s_0^2 \geq \tilde{v}(0)$, then $(\tilde{b}, \tilde{v})$  solves \eqref{eq:proof_optimal_policy_relaxed}, and $(\tilde{b}, \tilde{v}) = (\hat{b}, \hat{v})$. If $\s_0^2 < \tilde{v}(0)$, then there is a unique time $t_0$ such that $\tilde{v}(t_0) = \s_0^2$. Since the time-$t_0$ obedience constraint is active, it follows from Bellman's principle of optimality that the map $t \mapsto (\tilde{b} (t_0 + t), \tilde{v} (t_0 + t))$, which equals $(\hat{b}, \hat{v})$, solves \eqref{eq:proof_optimal_policy_relaxed}. Otherwise, replacing the time-$t_0$ continuation policy in $(\tilde{b}, \tilde{v})$ with a solution of \eqref{eq:proof_optimal_policy_relaxed} would strictly increase the sender's payoff, while preserving all obedience constraints in \eqref{eq:proof_optimal_policy_aux}.

To prove that $(\tilde{b}, \tilde{v})$ solves \eqref{eq:proof_optimal_policy_aux}, attach nonnegative Lagrange multipliers $e^{-r t} \l(t)$ to each time-$t$ obedience constraint and $r e^{-r t} \mu(t)$ to each time-$t$ nonnegativity constraint. Integrate over these constraints to form the Lagrangian
\begin{align*}
L (b, v;\l, \mu)
&= \int_{0}^{\infty} r e^{-r t} \big[ (b(t) - \b)^2 + v(t) \big] \de t \\
& \quad + \int_{0}^{\infty} e^{-r t} \l(t)  \Set{ \int_{t}^{\infty} r e^{-r (s - t)} \big[ b^2(s) + v(s) \big] \de s - \frac{ \s^2 + r v(t)}{r - 2 \k} } \de t \\
& \quad  - \int_{0}^{\infty} r e^{-r t} \mu(t) v(t) \de t.
\end{align*}
After splitting the term in braces,%
\footnote{
	It suffices to define the Lagrangian for functions $(b,v)$ that yield finite loss for the sender. For such functions, both integrals are finite as long as $\l$ and $\mu$ are bounded, as they will be below. 
}
the double integral in the obedience constraint can be rearranged as
\[
	\int_{0}^{\infty} \int_{t}^{\infty} r e^{-r s} \l(t) \big[ b^2(s) + v(s) \big] \de s \de t
	= \int_{0}^{\infty} r e^{-r s}  \Paren{\int_{0}^{s} \l(t) \de t} \big[ b^2(s) + v(s) \big] \de s,
\]
where I have switched the order of integration by Tonelli's theorem. Next, swap the dummy variable names $s$ and $t$ in this integral, and define the accumulated multiplier
\[
\L(t) = \int_{0}^{t} \l(s) \de s.
\]

After these simplifications, we have
\begin{equation*}
	L(b, v;\l, \mu)  = \int_{0}^{\infty} r e^{- r t} \ell \big( b(t), v(t); \l(t), \mu(t) \big) \de t
	- \frac{ \s^2}{r - 2 \k} \int_{0}^{\infty} e^{-r t} \l(t) \de t,
\end{equation*}
where $\ell \big( b(t), v(t); \l(t), \mu(t) \big)$ equals
\begin{equation} \label{eq:proof_optimal_policy_integrand}
	(b(t) - \b)^2 + \L(t) b^2(t) + \big( 1 + \L(t) - \l(t)/(r - 2 \k) - \mu(t) \big) v(t).
\end{equation}
Define the nonnegative multipliers by 
\begin{equation*}
	\big( \hat{\l}(t), \hat{\mu}(t) \big)
	= \begin{cases}
	\big( (r - 2 \k) e^{(r - 2 \k) t}, 0 \big) 
	&\text{if}~ t < T,\\
	\big(0,e^{(r - 2 \k)T} \big) 
	&\text{if}~t \geq T.
	\end{cases}
\end{equation*}
With these multipliers, the coefficient on $v(t)$ vanishes, and the integrand in \eqref{eq:proof_optimal_policy_integrand} becomes
\[
	(b(t) - \b)^2 + \big( e^{(r -2 \k) (t \wedge T)} - 1\big) b^2(t).
\]
This expression is convex in $b(t)$ and the first-order condition gives
\[	
	b(t) = \b e^{- (r - 2 \k) (t \wedge T )} = \frac{\s}{\sqrt{r - 2 \k}} e^{(r - 2\k) (T - t)_+} = \tilde{b}(t),
\]
where the middle equality uses the identity $t \wedge T = T -( T - t)_+$ and the definition of $T$ in \eqref{eq:modified_T}. 

It follows that $(\tilde{b}(t), \tilde{v}(t))$ minimizes $\ell ( \cdot, \cdot; \hat{\l}(t), \hat{\mu}(t))$  for each time $t$, hence $(\tilde{b}, \tilde{v})$ minimizes $L( \cdot, \cdot; \hat{\l}, \hat{\mu})$. It can be checked that $(\tilde{b}, \tilde{v})$ satisfies feasibility and complementary slackness. Therefore, all the Kuhn--Tucker conditions are satisfied.

\paragraph{Uniqueness} I claim that if a function $(b,v)$ solves \eqref{eq:proof_optimal_policy_relaxed}, then $(b(t), v(t)) = (\hat{b}(t), \hat{v}(t))$ for almost every time $t$. To see this, suppose that $(b,v)$ solves \eqref{eq:proof_optimal_policy_relaxed}. Then $b(t) = \hat{b}(t)$ for almost every $t$; otherwise, $( b/2 + \hat{b}/2, v/2 + \hat{v}/2)$ is a feasible strict improvement.  Furthermore, $v(t) = \hat{v}(t)$ for almost every $t$; otherwise, $(b, v \wedge \hat{v})$ is a feasible strict improvement, where $v \wedge \hat{v}$ denotes the pointwise minimum of $v$ and $\hat{v}$.\footnote{In fact, a slightly stronger result holds. Since $(\hat{b}, \hat{v})$ satisfies the obedience constraint with equality for every time $t$, it follows that $v(t) \geq \hat{v}(t)$ for every time $t$. If $(b,v)$ also satisfies the no-disclosure upper bounds, then $v$ must be lower semicontinuous, and hence  $v(t) = \hat{v}(t)$ for \emph{every} time $t$.}

\subsection{Proof of  \texorpdfstring{\cref{cor:optimal_deterministic}}{Corollary \ref{cor:optimal_deterministic}}}

For uniqueness, follow the argument from the proof of \cref{thm:optimal_policy} in \cref{subsec:proof_optimal_policy}. For optimality, observe that the sender's objective, denoted $u_S (b, v)$, is independent of the volatility parameter $\s$. In the relaxed problem (without the initial variance constraint) from \eqref{eq:proof_optimal_policy_aux}, the feasible set is increasing (with respect to set inclusion) in $\s$. For each $\s > 0$, let $(\tilde{b}_\s, \tilde{v}_\s)$ denote the solution of \eqref{eq:proof_optimal_policy_aux} when the volatility equals $\s$. Define $(\tilde{b}_0, \tilde{v}_0)$ by taking $b_0 = \b$ in the expressions from the statement of \cref{cor:optimal_deterministic}. Fix a positive sequence $\{ \s_n\}$ satisfying $\s_n \downarrow 0$. It suffices to check that $u_S ( \hat{b}_{\s_n}, \hat{v}_{\s_n}) \to u_S (\hat{b}_0, \hat{v}_0)$. Observe that $(\tilde{b}_{\s_n}, \tilde{v}_{\s_n})$ converges pointwise to $(\tilde{b}_0, \tilde{v}_{0})$. By \cref{prop:CS}, $(\tilde{b}_{\s_n}(t) - \b)^2 + \tilde{v}_{\s_n} (t)$ is monotonically increasing in $n$ for each $t$, so by Lebesgue's monotone convergence theorem, $u_S ( \tilde{b}_{\s_n}, \tilde{v}_{\s_n}) \to u_S (\tilde{b}_0, \tilde{v}_0)$. Therefore, $(\tilde{b}_0, \tilde{v}_0)$  is optimal in  \eqref{eq:proof_optimal_policy_aux}. By Bellman's principle of optimality,  $(\hat{b}_0, \hat{v}_0)$ is optimal in \eqref{eq:proof_optimal_policy_relaxed}.

\subsection{Proof of \texorpdfstring{\cref{prop:CS}}{Proposition \ref{prop:CS}}} \label{subsec:proof_comparative_statics}
Assume that $\b > \s / \sqrt{r - 2 \k}$ and the initial variance constraint is not active. We have
\begin{equation*}
	\hat{b}(t) = \max \Set{ \b e^{- (r -2 \k) t}, \frac{\s}{\sqrt{r -2 \k}}},
\end{equation*}
so the comparative statics for $\hat{b}$ are clear. For the variance, observe that the expression for $\hat{v}$ in the theorem statement is strictly increasing as a function of $T$ and it is otherwise independent of $\b$. Since the full-disclosure time $T$ is strictly increasing in $\b$, it follows that $\hat{v}$ is increasing in $\b$. 

Finally, I check that $\hat{v}$ is decreasing in $\s$. Fix $\s_1$ and $\s_2$ with $ 0 < \s_1 < \s_2$. For each volatility parameter $\s_i$, denote the optimal bias--variance pair by $(\hat{b}_i, \hat{v}_i)$ and the full-disclosure time by $T_i$. We have $T_1 > T_2$. For $t \geq T_1$, we have $\hat{v}_1(t) = \hat{v}_2(t) = 0$. For $T_2 \leq t < T_1$, we have $\hat{v}_1 ( t) > \hat{v}_2(t) = 0$. Finally, for $t < T_2$, observe that $\hat{b}_i(t)$ is independent of $i$ and 
\[
	(r - 2\k) \hat{b}_i^2(t) = 2 \k \hat{v}_i(t) + \s_i^2 - \hat{v}_i'(t),
\]
hence
\[
	\hat{v}_1'(t) - \hat{v}_2'(t) = 2 \k ( \hat{v}_1(t) - \hat{v}_2(t)) + \s_1^2 - \s_2^2 < 2 \k ( \hat{v}_1(t) - \hat{v}_2(t)).
\]
Over the interval $(0, T_2)$, the function $f(s) =  \hat{v}_2 (T_2 - s) - \hat{v}_1 ( T_2 - s)$ satisfies $f'(s) \leq - 2 \k f(s)$.  By Gr\"{o}nwall's inequality, $f(s) \leq e^{-2 \k  s } f(0)$ for all $s$ in $[0, T_2]$, hence
\[
	\hat{v}_1(t) - \hat{v}_2(t) \geq e^{- 2 \k} ( \hat{v}_1(T_2) - \hat{v}_2(T_2)) > 0,
\]
for all $t$ in $[0, T_2]$. 

\subsection{Proof of \texorpdfstring{\cref{thm:multidim}}{Theorem \ref{thm:multidim}}} \label{subsec:proof_multidim}

Assume  $\| \b \| > \hat{\s}$, for otherwise the result is clear. It can be shown that the full-disclosure times are well-defined; for details, see the last part of the proof.  Taking as given that these times are well-defined, I prove the result. 

Drop the no-disclosure bounds except those starting at time $0$ to obtain the relaxed problem
\begin{equation} \label{eq:proof_multidim_relaxed}
	\begin{aligned}
		& \text{minimize} && \int_{0}^{\infty} r e^{-r t} \Brac{ \|  b(t) - \b \|^2  + \sum\nolimits_{i=1}^{n} v_i(t) } \de t \\
		&\text{subject to} && \int_{t}^{\infty} r e^{-r(s - t)} \Brac{ \| b(s) \|^2 + \sum\nolimits_{i=1}^{n} v_i(s)} \de s \leq \sum_{i=1}^{n} \frac{ \s_i^2 + r v_i(t)}{r - 2 \k_i}, \quad t \geq 0 \\
		&&& 0 \leq v_i(t) \leq \h_i ( \s_{0,i}^2, t), \quad i = 1, \ldots, n, \quad t \geq 0.
	\end{aligned}
\end{equation}
It is straightforward to check that $\hat{v}$ satisfies the dropped no-disclosure bounds. 
I prove that $(\hat{b}, \hat{v})$ solves \eqref{eq:proof_multidim_relaxed}, and then I check uniqueness. 

\paragraph{Optimality} To handle the initial variance constraint, consider an auxiliary problem. If $i_0 = 1$ and $\| \hat{b}(0)\| < \| \b\|$, define $t_{0}$ so that $\hat{b}(t_0) = \b$ (where we define $\hat{b}$ and $\hat{v}$ on all of $\R$ by the expressions in the theorem statement). Otherwise, set $t_0 = 0$. The auxiliary problem is to choose functions $b$ and $v$ on $[t_0, \infty)$ to solve
\begin{equation} \label{eq:proof_multidim_aux}
	\begin{aligned}
		& \text{minimize} && \int_{t_{0}}^{\infty} r e^{-r t} \Brac{ \| b(t) - b\|^2  + \sum\nolimits_{i=1}^{n} v_i(t) } \de t \\
		&\text{subject to} && \int_{t}^{\infty} r e^{-r(s - t)} \Brac{ \| b(s)\|^2 + \sum\nolimits_{i=1}^{n} v_i(s)} \de s \leq \sum_{i=1}^{n} \frac{ \s_i^2 + r v_i(t)}{r - 2 \k_i}, \quad t \geq t_{0} \\
		&&& v_i(t) \geq 0, \quad i = 1, \ldots, n, \quad t \geq t_{0}\\
		&&& v_i(t) \leq \h_i ( \hat{v}_i(t_0), t - t_0), \quad i = 1, \ldots, n, \quad t\geq t_0.
	\end{aligned}
\end{equation}

Define auxiliary functions $\tilde{b}$ and $\tilde{v}$ on $[t_0, \infty)$ by the expressions for $\hat{b}$ and $\hat{v}$ (with the new definition of $t_0$). Below, I will show that $(\tilde{b}, \tilde{v})$ solves \eqref{eq:proof_multidim_aux}. I claim that this implies that $(\hat{b}, \hat{v})$ solves \eqref{eq:proof_multidim_relaxed}. If $t_0 = 0$, then $(\tilde{b}, \tilde{v})$ solves \eqref{eq:proof_multidim_relaxed}, and $(\hat{b}, \hat{v})$ equals $(\hat{b}, \hat{v})$. If $t_0 < 0$, then $\tilde{v}_i (0) = \s_{0,i}^2$ for all $i = 1, \ldots, n$. Since the time-$0$ obedience constraint is active, it follows from Bellman's principal of optimality that the restriction of $(\tilde{b}, \tilde{v})$ to $[0, \infty)$, which equals $(\hat{b}, \hat{v})$, solves \eqref{eq:proof_multidim_relaxed}. Otherwise, replacing the time-$0$ continuation policy in $(\tilde{b}, \tilde{v})$ with a solution of \eqref{eq:proof_multidim_relaxed} would strictly increase the sender's payoff, while preserving all obedience constraints in \eqref{eq:proof_multidim_aux}. 

To prove that $(\tilde{b}, \tilde{v})$ solves \eqref{eq:proof_multidim_aux}, attach nonnegative Lagrange multipliers $e^{-r t} \l(t)$ to each time-$t$ obedience constraint, $r e^{-r t} \mu_i(t)$ to the nonnegativity constraint on $v_i(t)$, and $r e^{-r t} \g_i(t)$ to the no-disclosure bound on $v_i(t)$. Integrate over these constraints to form the Lagrangian $L (b, v, \l, \mu, \g)$. Simplifying as in the single-dimensional case, we have
	\begin{align*}
			L(b,v; \l, \mu, \g)& = \int_{t_0}^{\infty} r e^{-r t} \ell \big( b(t), v(t), \l(t), \mu(t), \g(t)\big) \de t  \\
		&- \sum_{i=1}^{n} \frac{\s_i^2}{r - 2\k_i} \int_{t_0}^{\infty} e^{-r t} \l(t) \de t\\
		&- \sum_{i =1}^{n} \int_{t_0}^{\infty} r e^{-r t} \g_i(t) \h_i ( \hat{v}_i(t_0), t - t_0) \de t,
	\end{align*}
	where the integrand $\ell \big( b(t), v(t), \l(t), \mu(t) , \g(t)\big)$ equals 
	\begin{equation} \label{eq:proof_multidim_integrand}
		\| b(t)  - \b\|^2  + \L(t)  \| b(t)\|^2 + \sum_{i=1}^{n} \Paren{1 + \L(t) - \frac{\l(t)}{r - 2\k_i} - \mu_i(t) + \g_i(t )} v_i(t).
	\end{equation}
	Next, define the multipliers as follows. First, define $j \colon [t_0, t_n) \to \{i_0, \ldots, n\}$ by $j(t) = i$ if $t \in [t_{i-1}, t_i)$. Let 
	\[
	S(t)
	= \int_{t_0}^{t} (r - 2 \k_{j(s)}) \de s
	=  \sum_{i=i_0}^{n} (r - 2 \k_i) (t \wedge t_i - t_{i-1})_+.
	\]
Let
	\[
	\hat{\l}(t) 
	= \begin{cases} 
		\big( r - 2 \k_{j(t)} \big) e^{S(t)}
		&\text{if}~t < t_n\\
		0
		&\text{if}~ t \geq t_n.
	\end{cases}
	\]	
	It can be shown that that $\hat{\L}(t) = e^{S(t)} - 1$. Next, for $i = 1, \ldots, n$, let 
	\[
		( \hat{\mu}_i (t), \hat{\g}_i(t)) 
		=
		\begin{cases}
		\bigl(0, 	2 ( \k_i - \k_{j(t)})(r - 2 \k_i)^{-1} e^{S(t)} \bigr)	&\text{if}~t < t_{i-1}, \\
		(0,0)	&\text{if}~t_{i-1} \leq t < t_{i}, \\
		\bigl(	2 (\k_{j(t)} - \k_i )(r - 2 \k_i)^{-1} e^{S(t)}, 0 \bigr)	&\text{if}~t_i \leq t.
		\end{cases}
	\]

	In \eqref{eq:proof_multidim_integrand}, the coefficients on each $v_i(t)$ vanish, leaving
	\[
	\| b(t) -  \b\|^2 + (e^{S(t)} - 1) \| b (t) \|^2.
	\]
	This expression is convex in $b(t)$ and the first-order condition gives $	b(t) = \b \exp (- S(t) )$. From the definition of $t_0$, we have 
	\[
		\b = \hat{b}(t_0) = \hat{\s} \exp   \Paren{ S(t_n) } \frac{\b}{\| \b\|}.
	\]
	Combining these identities, we have
	\[
		b(t) = \hat{\s} \exp \Paren{ S(t_n) - S(t) } \frac{\b}{\| \b \|} =  \tilde{b}(t).
	\]
	
	It follows that $(\tilde{b}(t), \tilde{v}(t))$ minimizes $\ell ( \cdot, \cdot; \hat{\l}(t), \hat{\mu}(t), \hat{\gamma}(t))$ for each time $t$, hence $(\tilde{b}, \tilde{v})$ minimizes $L ( \cdot, \cdot; \hat{\l}, \hat{\mu}, \hat{\g})$. It can be checked that $(\tilde{b}, \tilde{v})$ satisfies feasibility and complementary slackness. Therefore, all the Kuhn--Tucker conditions are satisfied.

	\paragraph{Uniqueness} Here, the argument is different than in the single-dimensional case. By Bellman's principle optimality, it suffices to show that $(\tilde{b}, \tilde{v})$ is the (almost everywhere) unique solution of \eqref{eq:proof_multidim_aux}. If a function $(b,v)$ solves $\eqref{eq:proof_multidim_aux}$, then $L( b, v;  \hat{\l}, \hat{\mu}, \hat{\g}) = L ( \tilde{b}, \tilde{v} ; \hat{\l}, \hat{\mu}, \hat{\g})$ and $(b,v)$ satisfies complementary slackness with the multipliers $(\hat{\l}, \hat{\mu}, \hat{\g})$. For each $t$, 
	\[
		\argmax_{(x,y) \in \R \times \R_+} \ell( x, y;  \hat{\l}(t), \hat{\mu}(t), \hat{\g}(t)) = \{ \hat{b}(t)\} \times \R_+.
	\]
	Therefore, $b(t) = \hat{b}(t)$ for almost all $t$. For each $t$, we have  $\hat{\l} (t) > 0$. If $\k_1 < \cdots < \k_n$, then for each $i$, we have $\hat{\g}_i(t) > 0$ for $t < t_{i-1}$ and $\hat{\mu}_i (t) > 0$ for $t \geq t_i$. Therefore, complementary slackness implies that $v(t) = \hat{v}(t)$ for almost all $t$.\footnote{In fact, this equality must hold for every $t$, by the same argument as in the single-dimensional case. If the $\k_i$ agree for  $i$ in some subinterval $I$ of $\{1, \ldots, n\}$, then $\hat{\mu}_i(t) = \hat{\g}_i(t) = 0$ for all $t$ in $T = \cup_{i \in I} [t_{i-1}, t_i)$. Only the sum $\sum_{i \in I} v_i(t)$ is pinned down for $t$ in $T$. It is optimal to select any Bayes-plausible choices of $v_i(t)$, for $i$ in $I$ and $t$ in $T$, that induce the correct sum.}

\paragraph{Definition of full-disclosure times} Suppose that $\s_{0,i}^2 > 0$ for some $i$, for otherwise the full-disclosure times are all $0$. 

In the theorem statement, the bias and variance functions are expressed in terms of the future disclosure times. To prove that these full-disclosure times are well-defined, I express the bias and variance functions in terms of their values at time $0$. The formal procedure follows. 

Let $\VV$ contain the zero $n$-vector together with all $n$-vectors of the form $(0_{i_0 - 1}, \nu_{i_0}, \s_{0, i_0 + 1}^2, \ldots, \s_{0,n}^2)$, for some component $i_0 \in \{1, \ldots, n\}$ and some $\nu_{i_0}$ in $(0, \s_{0,i}^2]$, where $0_{i_0 -1}$ denotes a zero vector with $i_0  -1$ components. The set $\VV$ is totally ordered by the usual componentwise order. Fix $\a \in [ \hat{\s}, \infty)$ and $\nu \in \VV$. Define functions $b$ and $v$ and disclosure times $t_1, \ldots, t_n$ as follows. Set $t_i =0$ for $i \leq i_0  - 1$. Let $v(0) = \nu$  and $b(0) = \a \b / \| \b\|$. For each $i$, given that $b(t)$ and $v(t)$ are defined for $t \leq t_{i-1}$, define the time $t_i$ and the values $(b(t),v(t))$ for $t$ in $(t_{i-1}, t_i]$ as follows. For the bias,
\[
	b (t) = b(t_{i-1})  e^{-( r- 2 \k_i) (t - t_{i-1})}.
\]
For $j \leq i -1$, we have $v_j(t) = 0$. For $j \geq i + 1$, we have $v_j(t) = \h  ( v_j ( t _{i-1}), t - t_{i-1})$. Finally, $v_i (t)$ can be expressed explicitly, but it is more convenient to observe that it is the unique solution of the differential equation
\[ 
	v_i'(t) =  2 \k_i v_i (t)   - (r - 2 \k_i)  ( \| b (t) \|^2 - \hat{\s}_i^2),
\]
with the given boundary value $v_i ( t_{i-1})$. Let $t_{i}$ be the smallest time $t$ such that either $v_i (t) = 0$ or $ \| b(t) \| = \hat{\s}$. If  $ \| b (t_i) \| = \hat{\s}$, set $T = t_i$ and terminate the procedure. If $ \| b(t_i) \| > \hat{\s}$, either proceed to the next step if $i < n$ or else set $T = t_n$ if $i = n$. This procedure determines a function $f \colon [ \hat{\s}, \infty) \times \VV  \to \R$ by setting
\[
	f (\a, \nu) =  \| b(T) \| - \hat{\s} - (v_1(T) + \cdots + v_n (T) ).
\]

This function $f$ is continuous in the parameters of the problem. I claim that $f$ satisfies the following monotonicity properties:
\begin{enumerate}[label = (\roman*), ref = \roman*]
	\item \label{it:SC_in_b} For each $\nu \in \VV$, the function $f(\cdot, \nu)$ is strictly single-crossing from below.
	\item \label{it:SC_in_v} For each $\a \in [\hat{\s}, \infty)$, the function $f(\a, \cdot)$ is strictly single-crossing from above. 
\end{enumerate}
First I complete the proof, taking these properties as given. Fix $\nu \in \VV$. We have $f(\hat{\s}, \nu) = - (\nu_1 + \cdots + \nu_n) \leq 0$. It can be shown that for $\a$ sufficiently large, $f(\a, \nu) > 0$. By continuity and \eqref{it:SC_in_b}, there is a unique value $\a^\ast ( \nu) \in [\hat{\s}, \infty)$ such that $f( \a ^\ast (\nu), \nu) = 0$. By \eqref{it:SC_in_b} and \eqref{it:SC_in_v},  the function $\a^\ast \colon \VV\to [ \hat{\s}, \infty)$ is strictly increasing. Observe that $\a^\ast (0) = \hat{\s}$. Let $\hat{\a} = \a^\ast ( \s_{0,1}^2, \ldots, \s_{0,n}^2)$. By continuity, the image $\a^\ast (\VV)$ is the interval $[\hat{\s}, \hat{\a}]$, so $\a^\ast$ has a right inverse $(\a^\ast)^{-1} \colon [\hat{\s}, \hat{\a}] \to \VV$. Therefore, the full-disclosure times are pinned down by applying the procedure above with the initial conditions $\hat{b}(0) = ( \| \b \| \wedge \hat{\a})  \b/ \| \b\|$ and $\hat{v}(0 )  = (\a^\ast)^{-1} ( \| \b \| \wedge \hat{\a})$.

To prove \eqref{it:SC_in_b}, fix some nonzero $\nu \in \VV$. (The result is clear with $\nu = 0$ since $f(\a, 0)  = \a - \hat{\s}$ for all $\a$.) Let $i_0$ denote the index of the first nonzero component of $\nu$. Fix $\a, \bar{\a} \geq \hat{\s}$ with $\bar{\a} > \a$. Apply the  procedure above from the initial conditions $(\a, \nu)$ and $(\bar{\a}, \nu)$ to obtain $(b, v, t_1, \ldots, t_n, T)$ and $(\bar{b}, \bar{v}, \bar{t}_1, \ldots, \bar{t}_n, \bar{T})$, respectively. Suppose $f(\a, \nu) \geq 0$. Hence, $v(T) = 0$.  I show that $f(\bar{\a}, \nu) > 0$. Since $\k_1 \leq \cdots \leq \k_n$, it can be shown using Gr\"{o}nwall's inequality that:
\begin{enumerate}[label = (\alph*)]
	\item $\bar{t}_i < t_i$ for  $i \geq i_0$;
	\item  $\| \bar{b} (t)\| > \| b(t) \|$ for $t \leq \min \{ \bar{T}, T\}$; 
	\item $\bar{v}_i(t) \leq v_i(t)$ for $t \leq \min \{ \bar{T}, T\}$.
\end{enumerate}
 In particular, we cannot have $\bar{T} > T$, for then $ \bar{v}( T)  \leq  v(T)  = 0$. So $\bar{T} \leq T$, and hence $\| \bar{b} (\bar{T}) \| >  \| b (T)\| \geq \hat{\s}$. Thus, $f(\bar{\a}, \nu) > 0$. 

To prove \eqref{it:SC_in_v}, fix $\a > \hat{\s}$. (The result is clear with $\a = \hat{\s}$ since $f(\hat{\s}, \nu) = - (\nu_1 + \cdots + \nu_n)$ for all $\nu$.) Fix $\nu, \bar{\nu} \in \VV$ with $\bar{\nu} > \nu$. Apply the  procedure above from the initial conditions $(\a, \nu)$ and $(\a, \bar{\nu})$ to obtain $(b, v, t_1, \ldots, t_n, T)$ and $(\bar{b}, \bar{v}, \bar{t}_1, \ldots, \bar{t}_n, \bar{T})$, respectively. Suppose $f( \a, \nu ) \leq 0$.  Hence, $\| b(T)\| = \hat{\s}$, which implies that $\nu$ is nonzero. I show that $f(\a, \bar{\nu}) < 0$. Since $\k_1 \leq \cdots \leq \k_n$, it can be shown using Gr\"{o}nwall's inequality that:
\begin{enumerate}[label = (\alph*)]
	\item for $i \geq i_0 (a, \bar{\nu})$, if $\bar{v}_i (\bar{T}) = 0$, then $\bar{t}_i > t_i$;
	\item $\| \bar{b} (t)\|\leq \| b(t) \|$ for $t \leq \min \{ \bar{T}, T\}$; 
	\item $\bar{v}_i(t) \geq v_i(t)$ for $t \leq \min \{ \bar{T}, T \}$.
\end{enumerate}
 In particular, we cannot have $\bar{T} > T$, for then $ \| \bar{b} (T) \| \leq \| b(T) \| = \hat{\s}$. So $\bar{T} \leq T$, and hence $\bar{v}_n (\bar{T})  > 0$, for otherwise $t_n < \bar{T} \leq T$, which is a contradiction. Thus, $f (\a, \bar{\nu}) < 0$. 

\newpage

\bibliographystyle{ecta}
\bibliography{bibliography_DID}

\end{document}